\documentclass[preprint,12pt]{elsarticle}

\usepackage{xspace}
\usepackage{orcidlink}
\usepackage{amsmath,amssymb,bm}
\usepackage{graphicx}
\usepackage{booktabs}
\usepackage{siunitx}
\usepackage{hyperref}
\usepackage{algorithm}
\usepackage{algpseudocode}
\usepackage{listings}
\usepackage{pdfcomment}
\definecolor{codegray}{gray}{0.95}
\lstdefinestyle{cli}{
  backgroundcolor=\color{codegray},
  basicstyle=\ttfamily\small,
  breaklines=true,
  frame=none,
  columns=fullflexible,
  showstringspaces=false,
  xleftmargin=1em,
  framexleftmargin=1em,
}

\newcommand{\adda}{\texttt{ADDA}\xspace}
\newcommand{\hlc}[3][1]{%
    \ifcase#1\relax
    \or
        \pdfmarkupcomment[markup=Highlight,color=yellow,author=Clément]{#2}{#3}%
    \or
        \pdfmarkupcomment[markup=Highlight,color=cyan,author=Maxim]{#2}{#3}%
    \fi
}
\DeclareMathOperator{\Impart}{Im}
\DeclareMathOperator{\Repart}{Re}
\DeclareSIUnit[]{\GiB}{\gibi\byte}

\begin{document}
\sloppy %Allow LaTeX to hyphenate or break links

\begin{frontmatter}

\title{Efficient refractive-index sweeps in the discrete dipole approximation using shifted Krylov methods}

\author[1]{Clément Argentin\,\orcidlink{0009-0008-1270-2428}}
\author[2]{Konstantin G. Inzhevatkin\,\orcidlink{0000-0003-2858-2447}}
\author[1]{Maxim A. Yurkin\,\orcidlink{0000-0002-3524-0093}\corref{cor1}}
\ead{yurkin@gmail.com}
\cortext[cor1]{Corresponding author.}

\affiliation[1]{organization={Université Rouen Normandie, INSA Rouen Normandie, CNRS}, addressline={CORIA UMR 6614}, city={Rouen}, country={France}}
\affiliation[2]{organization={JSC Upstream Solutions}, city={Novosibirsk}, country={Russia}}

%-------------------------------------------------------------------------------
\begin{abstract}
The discrete dipole approximation (DDA) is a flexible numerical method for simulating electromagnetic scattering by particles of arbitrary shape and internal structure. In many applications, such as optical-database generation, the same particle geometry must be simulated for many values of the complex refractive index, which are currently handled independently with proportional increase of computational time. For homogeneous particles, however, variations of the refractive index affect only the diagonal polarizability term of the interaction matrix. The corresponding linear systems therefore differ only by scalar shifts of the identity matrix, making them suitable for shifted Krylov-subspace methods. In this work, we use a shifted Krylov solver adapted to the complex-symmetric formulation of the matrix and implement it in the \adda code. The method is evaluated for a homogeneous spherical particle over a two-dimensional domain of complex refractive indices. The shifted solver reproduces the extinction cross sections obtained from independent simulations with relative differences below the iterative-method tolerance, while reducing the number of matrix--vector products by a factor of 130 for the tested configuration. We further analyze how the achievable speedup depends on the number of refractive-index values, the computational backend, and the variation of the convergence rate across the parameter space. The final wall-time acceleration for large refractive-index sweeps was demonstrated to be up to 40 times, which can be further improved by optimizing the scattered-fields calculations that became a new bottleneck.
\end{abstract}
%-------------------------------------------------------------------------------
%-------------------------------------------------------------------------------
\end{frontmatter}
%-------------------------------------------------------------------------------
\section{Introduction}
%-------------------------------------------------------------------------------

Since analytical solutions of Maxwell's equations are limited to simple geometries, numerical methods are generally required to describe scattering by particles with irregular shapes or complex internal structures~\cite{WRIEDT2009,KAHNERT2016}. A wide range of computational approaches has therefore been developed, differing in their representation of the scatterer, numerical formulation, and range of applicability.

Among these methods, the discrete dipole approximation (DDA) is particularly attractive because of its conceptual simplicity and geometrical flexibility~\cite{PURCELL1973,DRAINE1988,YURKIN2023}. The scatterer is discretized into a collection of polarizable volume elements (voxels), usually arranged on a regular cubical lattice, whose electromagnetic interactions are described through the dyadic Green tensor. Since only the scatterer volume needs to be discretized, DDA can treat particles of arbitrary shape as well as inhomogeneous and anisotropic materials. Once the voxel polarizations have been determined, internal and near fields, integrated cross-sections, and angle-resolved scattering quantities can be evaluated.

This flexibility comes with a computational cost. A target containing $N$ voxels leads to a dense complex linear system with $n=3N$ unknowns. Direct inversion and explicit storage of the $n \times n$ interaction matrix are therefore impractical for most relevant discretizations. Instead, the system is generally solved using iterative methods, with the drawback that the number of iterations required to reach a prescribed residual tolerance is not known a priori and may become very large, or convergence may even fail for poorly conditioned systems.

Over the years, numerous computational improvements have been developed to reduce the cost of DDA simulations. These developments can be broadly divided into two categories: methods that reduce the computational cost of an individual simulation, and methods that reduce the cost of performing multiple simulations ~\cite{YURKIN2023}.

The first category includes methods that operate at different stages of the numerical solution. In a direct implementation, each matrix--vector product (MVP) scales as $\mathcal{O}(n^2)$. However, for voxels arranged on a regular lattice, Goodman et al.~\cite{GOODMAN1991} showed that the interaction can be expressed as a discrete convolution and evaluated using fast Fourier transforms (FFTs), reducing the cost of an MVP to $\mathcal{O}(N_{\mathrm{box}}\log N_{\mathrm{box}})$, where $N_{\mathrm{box}}=n_xn_yn_z$ denotes the number of grid cells in the computational box. Since one or more MVPs are required at each iteration, the number of iterations is thus a major determinant of the total solution time. The original implementation of Purcell and Pennypacker~\cite{PURCELL1973} employed a relatively simple iterative scheme, whereas Draine~\cite{DRAINE1988} introduced a conjugate-gradient-based method to accelerate convergence. Modern DDA implementations now employ a variety of Krylov-subspace methods~\cite{YURKIN2007,CHAUMET2024,SOGABE2022}, while further reductions in the number of iterations can be achieved through preconditioning~\cite{GROTH2020,CHAUMET2023,LANIER2026} and physically motivated initial guesses~\cite{INZHEVATKIN2022,CHAUMET2023}. The DDA formulation and discretization have also been improved to reduce the computational cost. For particles with large aspect ratios, non-cubic lattices based on rectangular parallelepipeds can improve the geometrical representation~\cite{SMUNEV2015}. Improved methods to calculate interaction of finite voxels~\cite{YURKIN2010,CHAUMET2004}, weighted-discretization schemes~\cite{PILLER1997}, and extrapolation techniques~\cite{YURKIN2006} can increase the accuracy obtained for a given discretization and thereby reduce the number of voxels required to reach a target accuracy.

Additional gains have also been obtained through optimized orientation-averaging schemes~\cite{PENTTILA2011}, shared- and distributed-memory parallelization~\cite{HOEKSTRA1998}, and GPU implementations~\cite{HUNTEMANN2011}. Most of these developments primarily reduce the cost of an individual DDA calculation.

Many applications, however, require large sets of related simulations rather than a single run. Optical-property databases, for example, involve variations of particle size, shape, orientation, wavelength, and complex refractive index. This motivates methods that reuse information common to a family of scattering problems instead of solving every case independently. A well-established example is the T-matrix formalism introduced by Waterman~\cite{WATERMAN1965,WATERMAN1971,MISHCHENKO2020}. Once the T-matrix has been computed for a particle at a fixed frequency and material state, responses to different incident fields can be obtained without repeating the full solution. Within the T-matrix framework, the shape-matrix (Sh-matrix) approach~\cite{PETROV2006} goes further by separating morphology-dependent quantities from the dependence on particle size and refractive index, allowing precomputed geometrical information to be reused across a family of particle parameters.

Several other related ideas have also been explored. In the simplest case, a previously computed solution, or an interpolation of several solutions, can be used as the initial guess for a nearby problem~\cite{MUINONEN2006,OKADA2008,UM2013}, providing a so-called ``warm start''. More general subspace-recycling techniques retain selected spectral or Krylov information and incorporate it into subsequent solution spaces through augmentation or deflation~\cite{PARKS2006,SOODHALTER2020}. When several right-hand sides corresponding to the same coefficient matrix are available simultaneously, block Krylov methods construct a common search space from multiple residual vectors~\cite{OLEARY1980,MOTEKI2026}. These techniques can reduce the cost of related simulations, although the attainable gain depends on the structure and similarity of the corresponding systems.

A stronger relationship appears when only the refractive index of a homogeneous and isotropic particle is varied. In this case, the interaction matrices differ only by a scalar multiple of the identity and therefore form a family of shifted linear systems~\cite{FREUND1993,SOGABE2022}. Such systems share the same Krylov subspace, allowing all refractive indices to be treated through a single Krylov process. The MVPs are therefore shared (computed once per iteration) among all shifted systems, while the additional systems are advanced through shift-dependent vector and scalar updates. This shift invariance distinguishes shifted Krylov methods from warm-start and general subspace-recycling approaches. The shift property of the volume-integral equation (on which DDA is based~\cite{YURKIN2007}) has been noticed by Budko and Remis~\cite{BUDKO2004}, who used it to construct reduced-order models based on the full basis of the Krylov subspace. Shifted Krylov methods are, however,  more flexible, performing a sweep in a single run without any prior assumptions on the iterative convergence (size of Krylov subspace). This makes it particularly attractive for constructing optical-property databases covering large regions of the complex refractive-index plane. 

In this work, we implement a shifted Krylov solver for refractive-index sweeps in the open-source \adda code~\cite{YURKIN2011}. Since the interaction matrix is complex symmetric rather than Hermitian, the proposed method is based on the biconjugate gradient (BiCG)~\cite{FLETCHER1976} for complex-symmetric systems, hereafter denoted BiCG$_{\mathrm{CS}}$~\cite{FREUND1992}. We then derive the shifted formulation of the system, and discuss its convergence, computational cost, and memory requirements. The implementation is validated against independent DDA simulations and evaluated using sequential CPU, distributed-memory, and OpenCL configurations. Finally, the complete execution time is analyzed by separating initialization, iterative solution, and scattering calculations, thereby identifying the achievable acceleration.

The remainder of this paper is organized as follows. Section~\ref{sec:dda} presents the DDA formulation. Section~\ref{sec:shifted} introduces shifted Krylov subspaces and derives the shifted BiCG$_{\mathrm{CS}}$ method. The numerical methodology is described in Section~\ref{sec:methodology}. Section~\ref{sec:results} presents and discusses the numerical results, and Section~\ref{sec:conclusion} summarizes the main conclusions and perspectives.

%-------------------------------------------------------------------------------
\section{The discrete dipole approximation}
\label{sec:dda}
%-------------------------------------------------------------------------------

DDA can be derived from the volume-integral formulation of Maxwell's equations~\cite{YURKIN2007,YURKIN2023}. For a nonmagnetic scatterer occupying a volume $V$, the electric field satisfies
\begin{equation}
\begin{aligned}
    \mathbf{E}(\mathbf{r})
    ={}&
    \mathbf{E}^{\mathrm{inc}}(\mathbf{r})
    +
    k^2\int_{V\setminus V_0}
    \overline{\mathbf{G}}(\mathbf{r},\mathbf{r}')
    \overline{\boldsymbol{\chi}}(\mathbf{r}')
    \mathbf{E}(\mathbf{r}')
    \,\mathrm{d}^3\mathbf{r}'
    \\
    &+
    \mathbf{M}(V_0,\mathbf{r})
    -
    \overline{\mathbf{L}}(\partial V_0,\mathbf{r})
    \overline{\boldsymbol{\chi}}(\mathbf{r})
    \mathbf{E}(\mathbf{r}),
\end{aligned}
\end{equation}
where $\mathbf{E}^{\mathrm{inc}}$ is the incident field, $k$ is the wavenumber, $\overline{\mathbf{G}}$ is the corresponding dyadic Green tensor, and $\overline{\boldsymbol{\chi}}$ is the electric-susceptibility tensor. The exclusion volume $V_0$ surrounding $\mathbf{r}$ is introduced to handle the singularity of the Green tensor. The depolarization dyadic $\overline{\mathbf{L}}$ accounts for the singular contribution, while $\mathbf{M}$ is a correction term associated with the finite size of $V_0$.

In DDA, the scatterer is discretized into $N$ voxels centered at positions $\mathbf{r}_i$, each of which is represented by an induced polarization $\mathbf{p}_i\in\mathbb{C}^3$. For each observation point, $V_0$ is set to a surrounding voxel, which polarizability is determined by values of  $\overline{\mathbf{L}}$ and $\mathbf{M}$. Finally, the volume integral is approximated by finite sum (interactions among the voxels) leading to
\begin{equation}
    \sum_{j=1}^{N}
    \mathbf{A}_{ij}\mathbf{p}_j
    =
    \mathbf{E}^\mathrm{inc}_i,
    \qquad i=1,\ldots,N,
\end{equation}
where the $3\times3$ blocks of the interaction matrix are given by
\begin{equation}
    \mathbf{A}_{ij}
    =
    \begin{cases}
        \overline{\boldsymbol{\alpha}}_i^{-1},
        & i=j,\\[2mm]
        -k^2 \overline{\mathbf{G}}_{ij},
        & i\neq j.
    \end{cases}
    \label{eq:dda_matrix_blocks}
\end{equation}
Here, $\overline{\boldsymbol{\alpha}}_i$ denotes the polarizability tensor of voxel $i$, while $\overline{\mathbf{G}}_{ij}$ describes the interaction between voxels $i$ and $j$ -- specific expressions for both of them depend on the chosen DDA formulation~\cite{ARGENTIN2026}. Using global vectors $\mathbf{p}$ and $\mathbf{b}$, the system can be written in the compact $3N$-dimensional form
\begin{equation}
    \mathbf{A}\mathbf{p}=\mathbf{b}.
    \label{eq:dda_system}
\end{equation}
The symmetry properties of this system follow directly from those of the Green tensor and the polarizability. For reciprocal media, the Green tensor satisfies the reciprocity relation
\begin{equation}
    \overline{\mathbf{G}}_{ij}
    =
    \overline{\mathbf{G}}_{ji}^{T}.
\end{equation}
Consequently, when the polarizability tensors are scalar or symmetric, the complete interaction matrix is complex symmetric rather than Hermitian,
\begin{equation}
    \mathbf{A}^{T}=\mathbf{A},
    \qquad
    \mathbf{A}^{H}\neq\mathbf{A}.
    \label{eq:complex_symmetry}
\end{equation}
This complex-symmetric structure is central to the choice of iterative methods considered in Section~\ref{sec:shifted}.

%-------------------------------------------------------------------------------
\section{Shifted complex-symmetric Krylov method}
\label{sec:shifted}
%-------------------------------------------------------------------------------
\subsection{Dependence on the refractive index}
\label{sec:dda_shift}
%-------------------------------------------------------------------------------

Consider a homogeneous and isotropic particle with complex refractive index $m$. All occupied (non-void) voxels then have the same scalar polarizability, $\overline{\boldsymbol{\alpha}}_i=\alpha(m)\overline{\mathbf{I}}$. According to Eq.~\eqref{eq:dda_matrix_blocks}, the dependence of the interaction matrix on the refractive index is therefore confined to its diagonal blocks, while the voxel--voxel interaction terms remain unchanged.

For each refractive index $m_s$, we define the corresponding scalar shift as
\begin{equation}
    \sigma_s
    =
    1/\alpha(m_s).
\end{equation}
The linear systems associated with the different refractive indices can then be expressed as the shifted family
\begin{equation}
    \left(
        \mathbf{A}_\mathrm{off}
        +
        \sigma_s\mathbf{I}
    \right)
    \mathbf{p}^{(s)}
    =
    \mathbf{b},
    \qquad
    s=1,\ldots,N_m ,
    \label{eq:dda_shifted_systems}
\end{equation}
where $\mathbf{A}_\mathrm{off}$ denotes the off-diagonal (refractive-index-independent) part of the interaction matrix and $N_m$ is the number of refractive indices. For fixed geometry, discretization, wavelength, and illumination, both $\mathbf{A}_\mathrm{off}$ and $\mathbf{b}$ are therefore common to all systems. The computational advantage of this representation follows from a fundamental property of Krylov subspaces, discussed in the next section.

Note that physical homogeneity and isotropy alone are not sufficient to guarantee the shifted structure of Eq.~\eqref{eq:dda_shifted_systems}. The DDA formulation must additionally assign the same scalar polarizability to every voxel. Tensorial or position-dependent polarizability corrections do not preserve this property. Examples include the corrected lattice dispersion relation (CLDR) polarizability prescription~\cite{GUTKOWICZ2004}, which may yield a diagonal polarizability tensor even for an isotropic material, and boundary corrections such as weighted discretization~\cite{PILLER1997}, which modify the effective polarizability of voxels near the particle boundary. Preconditioning may also destroy the scalar-shift structure~\cite{SOGABE2022}.

%-------------------------------------------------------------------------------
\subsection{Krylov-subspace shift invariance}
%-------------------------------------------------------------------------------

To establish this property, consider the linear system in Eq.~\eqref{eq:dda_system} with an initial approximation $\mathbf{p}_0$ and corresponding residual
\begin{equation}
    \mathbf{r}_0
    =
    \mathbf{b}
    -
    \mathbf{A}\mathbf{p}_0.
\end{equation}
The Krylov subspace of dimension $\ell$ generated by $\mathbf{A}$ and $\mathbf{r}_0$ is defined as
\begin{equation}
    \mathcal{K}_{\ell}(\mathbf{A},\mathbf{r}_0)
    =
    \operatorname{span}
    \left\{
        \mathbf{r}_0,
        \mathbf{A}\mathbf{r}_0,
        \ldots,
        \mathbf{A}^{\ell-1}\mathbf{r}_0
    \right\}.
\end{equation}
Krylov-subspace iterative methods construct successive approximations to the solution in the affine space
\begin{equation}
    \mathbf{p}_{\ell}
    \in
    \mathbf{p}_0
    +
    \mathcal{K}_{\ell}(\mathbf{A},\mathbf{r}_0).
\end{equation}
The Krylov process requires MVPs together with scalar operations, inner products, and vector updates. This is particularly advantageous for DDA, since the interaction matrix does not need to be explicitly stored and the MVPs can be evaluated efficiently using FFTs.

More importantly for the family of systems in Eq.~\eqref{eq:dda_shifted_systems}, Krylov subspaces are invariant under scalar shifts of the matrix~\cite{JEGERLEHNER1996}. For a family of shifted systems, the same Krylov subspace can therefore be shared when their initial residuals are identical, or more generally collinear~\cite{SOGABE2022}. In the present implementation, the zero vector is used as the initial approximation for every refractive index. Since all systems in Eq.~\eqref{eq:dda_shifted_systems} share the same right-hand side, their initial residuals are identical, $\mathbf{r}_0=\mathbf{b}$. Hence,
\begin{equation}
    \mathcal{K}_{\ell}
    \left(
        \mathbf{A}_\mathrm{off}+\sigma_s\mathbf{I},
        \mathbf{r}_0
    \right)
    =
    \mathcal{K}_{\ell}
    \left(
        \mathbf{A}_\mathrm{off},
        \mathbf{r}_0
    \right),
    \qquad
    s=1,\ldots,N_m .
\end{equation}

Shift invariance alone, however, only establishes that the systems share the same search space. An iterative solver must additionally provide an efficient procedure to generate a basis of this space and to update the corresponding solutions. For the complex-symmetric interaction matrix considered here, see Eq.~\eqref{eq:complex_symmetry}, this is achieved through the complex-symmetric Lanczos process~\cite{SOGABE2022}.

%-------------------------------------------------------------------------------
\subsection{Shifted complex-symmetric Lanczos process}
\label{sec:shifted_lanczos}
%-------------------------------------------------------------------------------

For such matrix, the Lanczos process generates a Krylov basis through a three-term recurrence using the bilinear form $\mathbf{u}^{T}\mathbf{v}$, without complex conjugation~\cite{CRAVEN1969}. Combined with the corresponding projection condition, this leads to the conjugate-gradient-type method commonly referred to as the conjugate orthogonal conjugate gradient (COCG)~\cite{VORST1990} or BiCG$_{\mathrm{CS}}$~\cite{FREUND1992}. In the following, we use BiCG$_{\mathrm{CS}}$. 

This construction is the complex-symmetric analogue of the standard Lanczos/CG procedure for Hermitian positive-definite matrices~\cite{HESTENES1952}. Unlike standard CG, however, the bilinear form does not define a positive-definite inner product. Consequently, the usual CG convergence guarantees do not apply, and breakdown of the Lanczos recurrence may occur.

The BiCG$_{\mathrm{CS}}$ iteration can be expressed in two algebraically equivalent forms: the usual recursive formulation~\cite{FLETCHER1976}, in which the residual and search-direction vectors are updated directly, and a Lanczos formulation based on the three-term recurrence of the Krylov basis~\cite{FROMMER1999}. In exact arithmetic, both formulations generate the same sequence of iterates~\cite{FROMMER1999}. 

The shifted solver implemented here adopts the Lanczos representation, following the shifted Lanczos formulations of Algorithm~5.1 in Ref.~\cite{BIRK2015} and Algorithm~5 in Ref.~\cite{FROMMER1999}, specialized here to complex-symmetric matrices through the corresponding bilinear products. This differs from the recursive shifted BiCG$_{\mathrm{CS}}$ formulation presented, for example, in Algorithm~4.7 of Ref.~\cite{SOGABE2022}. Compared with the recursive seed-based formulation, the Lanczos representation introduces a slightly larger constant overhead, which can be quantified as follows.

In Algorithm~\ref{alg:shifted-bicg-cs}, a single complex-symmetric Lanczos recurrence generates the basis vectors and scalar coefficients common to all refractive indices. This recurrence requires two additional vectors of size $\mathcal{O}(N)$, and two additional vector operations of cost $\mathcal{O}(N)$ per iteration. Then, similarly to the recursive seed-based formulation, only the shift-dependent scalar quantities, the solution vector $\mathbf{x}_k^{(s)}$, and the search-direction vector $\mathbf{p}_k^{(s)}$ are updated for each active shift $\sigma_s$. The additional overhead associated with the Lanczos recurrence is independent of $N_m$ and therefore becomes negligible for large refractive-index sweeps compared with the $2N_m$ shift-dependent vector updates, whose total cost scales as $\mathcal{O}(NN_m)$ per iteration. Moreover, the cost of these additional Lanczos operations remains substantially smaller than that of the shared MVP. Note that here $\mathbf{x}_k^{(s)}$ corresponds to the polarization vector $\mathbf{p}^{(s)}$ introduced in Eq.~\eqref{eq:dda_shifted_systems}, whereas $\mathbf{p}_k^{(s)}$ denotes the internal search-direction vector of the iterative method (following the standard notation of CG literature~\cite{BARRETT1994,SAAD2003,SOGABE2022}).

\begin{algorithm}[!ht]
\small
\caption{Shifted BiCG$_{\mathrm{CS}}$ in Lanczos form}
\label{alg:shifted-bicg-cs}
\begin{algorithmic}

\State $\mathbf{x}_0^{(s)}\gets\mathbf{0},\quad
       \mathbf{p}_0^{(s)}\gets\mathbf{0},
       \qquad s=1,\ldots,N_m$
\State $\mathbf{r}_0\gets\mathbf{b},\quad
       \beta_0\gets\sqrt{\mathbf{r}_0^T\mathbf{r}_0},\quad
       \mathbf{v}_0\gets\mathbf{0},\quad
       \mathbf{v}_1\gets\mathbf{r}_0/\beta_0$
\State $\mathcal{S}\gets\{1,\ldots,N_m\}$
\For{$k=1,\ldots,K_{\max}$}

    \Statex \hspace{\algorithmicindent}// Shared Lanczos recurrence

    \State $\mathbf{w}\gets\mathbf{A}_\mathrm{off}\mathbf{v}_k$
    \State $\alpha_k\gets\mathbf{v}_k^T\mathbf{w}$

    \State $\widetilde{\mathbf{v}}_{k+1}
        \gets
        \mathbf{w}
        -\alpha_k\mathbf{v}_k
        -\beta_{k-1}\mathbf{v}_{k-1}$

    \State $\beta_k
        \gets
        \sqrt{
        \widetilde{\mathbf{v}}_{k+1}^T
        \widetilde{\mathbf{v}}_{k+1}}$

    \State $\mathbf{v}_{k+1}
        \gets
        \widetilde{\mathbf{v}}_{k+1}/\beta_k$

    \Statex \hspace{\algorithmicindent}// Shift-dependent updates

    \ForAll{$s\in\mathcal{S}$}

        \If{$k=1$}
            \State $\ell_k^{(s)}\gets0,\qquad
                   u_k^{(s)}\gets\beta_0$
        \Else
            \State $\ell_k^{(s)}
                \gets
                \beta_{k-1}/d_{k-1}^{(s)},\qquad
                u_k^{(s)}
                \gets
                -\ell_k^{(s)}u_{k-1}^{(s)}$
        \EndIf
    
        \State $d_k^{(s)}
            \gets
            \alpha_k+\sigma_s
            -\beta_{k-1}\ell_k^{(s)}$
    
        \State $\mathbf{p}_k^{(s)}
            \gets
            \mathbf{v}_k
            -\ell_k^{(s)}\mathbf{p}_{k-1}^{(s)}$
    
        \State $\gamma_k^{(s)}
            \gets
            u_k^{(s)}/d_k^{(s)}$
    
        \State $\mathbf{x}_k^{(s)}
            \gets
            \mathbf{x}_{k-1}^{(s)}
            +\gamma_k^{(s)}\mathbf{p}_k^{(s)}$
    
        \State $\widehat{\rho}_k^{(s)}
            \gets
            \dfrac{
            |\gamma_k^{(s)}|
            \|\widetilde{\mathbf{v}}_{k+1}\|_2}
            {\|\mathbf{b}\|_2}$
    
        \If{$\widehat{\rho}_k^{(s)}\leq\eta$}
            \State $\mathcal{S}
                \gets
                \mathcal{S}\setminus\{s\}$
        \EndIf
    
    \EndFor

    \If{$\mathcal{S}=\varnothing$}
        \State \textbf{break}
    \EndIf

\EndFor

\State \Return $\{\mathbf{x}^{(s)}\}_{s=1}^{N_m}$

\end{algorithmic}
\end{algorithm}

Convergence is then monitored independently for every active system. Accordingly, once a system satisfies the prescribed relative residual tolerance $\eta$, it is removed from the set of active shifts and its solution is no longer updated. The shared Lanczos process, however, continues until all systems have converged. Consequently, the total number of global iterations is determined by the slowest-converging system in the refractive-index sweep. If the costs of the shift-dependent updates and other operations are neglected, the computational time is proportional to the number of MVPs. If $K_s$ denotes the number of iterations required by system $s$ when solved independently, the ideal speedup obtained by neglecting all shift-dependent operations is

\begin{equation}
    S_{\mathrm{ideal}}
    =
    \frac{\sum_{s=1}^{N_m} K_s}
         {\max_s K_s},
    \label{eq:ideal_speedup}
\end{equation}
with an upper bound of $N_m$. Thus, the attainable gain depends directly on the variability of the convergence rates across the shifted systems. In finite-precision arithmetic the shifted and independent formulations may converge at slightly different iterations. We therefore also define the observed MVP-count reduction as

\begin{equation}
    S_{\mathrm{count}}
    =
    \frac{\sum_{s=1}^{N_m} K_s}
         {K_{\mathrm{shift}}},
    \label{eq:mvp_count_speedup}
\end{equation}
where $K_{\mathrm{shift}}$ is the total number of shared MVPs performed by the shifted calculation. However, we use the same notation $K_s$ for both cases, assuming that the choice is obvious from the context.

This computational acceleration comes at the cost of an increased memory requirement. The memory overhead is evaluated here with respect to the existing BiCG$_{\mathrm{CS}}$ implementation in \adda, which uses the recursive formulation rather than the Lanczos representation. Although the Lanczos formulation requires only two additional common vectors relative to the recursive formulation, the version of \adda that was used for all performance tests included four additional full-size vectors for the solver. However, during the latest optimizations we reduced this number to three, which is reflected in the following memory model. In addition, the two shift-dependent vectors, $\mathbf{x}_k^{(s)}$ and $\mathbf{p}_k^{(s)}$, are allocated for each of the $N_m$ shifted systems. Since each vector contains $3N$ complex double-precision values, the additional memory (in bytes)  with respect to a BiCG$_{\mathrm{CS}}$ calculation is

\begin{equation}
    \Delta M
    =
    48N\left[5+2(N_m-1)\right]\unit{\byte}
    =
    96(N_m+1.5)N\unit{\byte}.
\label{eq:delta_M_CPU}
\end{equation}

The corresponding relative memory requirement can be written as

\begin{equation}
    \frac{M_{\mathrm{SBiCG}}}
         {M_{\mathrm{BiCG}}}
    =
    1+\frac{N_m+1.5}{C},
\end{equation}
where $C$ depends on the target occupancy, the computational-grid dimensions, and, in parallel mode, the number of MPI processes. For example, for a sphere with a $128^3$ grid in sequential mode, $C=8.61$. A detailed derivation of the memory model is given in~\ref{app:memory_model}.

In practice, the \texttt{-prognosis} option available in \adda provides a direct estimate of the memory required by a given configuration before running the simulation, including shifted calculations for any number of refractive indices.
 
%-------------------------------------------------------------------------------
\subsection{Numerical equivalence}
\label{sec:numerical_equivalence}
%-------------------------------------------------------------------------------

For a fixed discretization, the shifted and independent formulations solve exactly the same linear systems. The highest level of agreement is therefore expected when the shifted solver is compared with an independent BiCG$_{\mathrm{CS}}$ calculation. In exact arithmetic, both formulations generate the same sequence of iterates and residuals~\cite{FROMMER1999}. Actual computations, however, are performed in finite-precision arithmetic. In \adda, double precision is used, corresponding to a unit round-off of order \num{e-16}. Each arithmetic operation performed during the iterative process may therefore introduce a small rounding error. The accumulation of these errors can then lead to differences between the iterates generated by the shifted and independent formulations. 

For sufficiently short and numerically stable Krylov sequences, these differences may remain close to the double-precision round-off level. This regime is hereafter referred to as machine-precision agreement~\cite{ARGENTIN2026}. As the Krylov sequence becomes longer, accumulated round-off errors can progressively reduce the agreement toward the convergence-tolerance scale, hereafter referred to as $\eta$-level agreement~\cite{ARGENTIN2026}. This is expected for increasing size and/or $\Repart m$, and for small or zero $\Impart m$.

A related finite-precision effect concerns the distinction between the recursively updated residual used by the iterative algorithm and the true residual associated with the computed solution. In exact arithmetic, these quantities coincide. In finite precision, however, the recursive updates accumulate rounding errors and a residual gap may develop between the estimated residual and the true residual~\cite{GREENBAUM1997}.

By contrast, when the shifted solver is compared with other Krylov methods, the iterates are not expected to coincide even in exact arithmetic, because different algorithms impose different projection or minimization conditions and therefore follow different convergence histories. Nevertheless, for the same discretized system and the same stopping criterion, the resulting solutions and derived optical quantities are still expected to agree at the $\eta$-level.

A lower level of agreement may appear when the discretization is not kept fixed between the shifted and independent calculations. Indeed, in a standard individual \adda simulation, the discretization may be selected according to the refractive index in order to satisfy the prescribed DDA default choice for voxel size. Consequently, different refractive indices may require different numbers of voxels and computational grids. In shifted mode, however, all refractive indices must be solved on the same discretized target to have the same interaction operator. The common value of \texttt{-dpl} (number of voxels per wavelength), or equivalently the common grid, must therefore be sufficiently fine to satisfy the discretization criterion for every refractive index in the sweep. Refractive indices, for which a coarser discretization may suffice in an independent calculation, are thus over-discretized in the shifted one.

This distinction is important when shifted calculations are compared with independent \adda runs. In this case, the two calculations do not necessarily solve the same discretized linear system, even though they represent the same scattering problem. Their agreement can therefore decrease from the $\eta$-level, to the grid-level agreement (see Table~2 of Ref.~\cite{ARGENTIN2026}), which is governed by discretization and shape errors and typically correspond to only a few matching digits. This over-discretization also increases the computational cost, but this increase is typically small compared with the computational saving obtained by sharing the Krylov process.

The agreement between shifted and independent calculations is further quantified through the relative difference in the extinction cross section,

\begin{equation}
    \eta_\mathrm{ext}
    =
    \frac{
        \left|
        C_{\mathrm{ext}}^{\mathrm{SBiCG}}
        -
        C_{\mathrm{ext}}^{\mathrm{BiCG}}
        \right|
    }{
        \left|
        C_{\mathrm{ext}}^{\mathrm{BiCG}}
        \right|
    }.
    \label{eq:cext_error}
\end{equation}
For convenience, $-\log_{10}\eta_\mathrm{ext}$ is further used to estimate the number of matching digits between the two calculations.

Importantly, machine-precision agreement is not required for the shifted solver to provide a numerically accurate solution. Rather, it is used here as a diagnostic of algebraic equivalence, implementation correctness, and numerical reproducibility between the shifted and independent formulations, as was recently demonstrated for DDA codes in general~\cite{ARGENTIN2026}. Thus, such exceptional agreement is highly informative when it occurs, but its absence does not indicate any practically-relevant loss of accuracy

%-------------------------------------------------------------------------------
\section{Methodology}
\label{sec:methodology}
%-------------------------------------------------------------------------------
\subsection{Implementation in \adda}
%-------------------------------------------------------------------------------

The shifted solver was implemented in the \texttt{feature/shifted-cg-v2} branch and specific commit\footnote{\url{https://github.com/Argencle/adda/tree/35deda91}} was used for all benchmark runs. However, the code development continued after all benchmarks were completed (partly, motivated by the obtained results) -- some of these improvements are mentioned below. This development branch will be integrated into a future version of \adda.

One can notice that the formulation introduced in Eq.~\eqref{eq:dda_matrix_blocks} is not the linear-system formulation used by \adda, but corresponds to the standard formulation used, for example, in \texttt{DDSCAT}~\cite{ARGENTIN2026}. The default \adda formulation, denoted as the symmetrized form in Ref.~\cite{ARGENTIN2026}, corresponds to a Jacobi-preconditioned form of the system. Its diagonal blocks are equal to the identity, while the refractive-index dependence is transferred to the interaction terms. Consequently, variations of the refractive index can no longer be represented by a scalar shift of the identity, preventing direct application of the shifted formulation derived in Section~\ref{sec:dda_shift}.

To enable the shifted solver, we therefore introduced the option \texttt{-linear\_form}, which allows switching between the standard and symmetrized formulations of the linear system. Although this option is primarily required here to expose the shifted structure of the standard formulation, it also provides an additional means of verifying the numerical equivalence of different DDA formulations and implementations~\cite{ARGENTIN2026}. More generally, it may also be useful for the development and assessment of preconditioners.

The shifted solver is selected in \adda through the option \texttt{-iter sbicg}. Refractive indices are supplied using the existing \texttt{-m} command-line option. Whereas multiple refractive indices normally describe different materials in inhomogeneous or anisotropic targets, with \texttt{-iter sbicg} they are interpreted as the set of refractive indices to be treated simultaneously. Since the shifted formulation considered here applies only to homogeneous and isotropic particles, inhomogeneous and anisotropic targets are disabled in this mode, avoiding any ambiguity between these two uses of the \texttt{-m} option.

The shifted solver also uses the standard linear formulation (\texttt{-linear\_form standard}) and a zero initial field (\texttt{-init\_field zero}) by default. The option \texttt{-store\_shifted\_convergence} was introduced to record the convergence information of each individual shifted system. Without this option, only the global convergence of the simultaneous calculation is reported, whereas the additional output provides the iteration count and final residual associated with each refractive index.

The maximum number of simultaneously treated refractive indices is currently controlled at compile time by the constant \texttt{MAX\_N\_SHIFTED} defined in \texttt{const.h}. This value can be increased when larger sweeps are required.

%-------------------------------------------------------------------------------
\subsection{Benchmark configuration}
\label{sec:benchmark_configuration}
%-------------------------------------------------------------------------------

The benchmark considers a homogeneous sphere with size parameter $20$. The complex refractive index is sampled over $\Repart m\in[1.3,1.8]$ and $\Impart m\in[0.1,0.6]$ with a step of $0.025$ in both directions, resulting in a $21\times21$ grid containing $N_m=441$ refractive indices. The FCD formulation~\cite{YURKIN2010} is used throughout. This relatively broad refractive-index domain was selected to produce a substantial variation in convergence behavior and thereby provide a demanding test of the shifted solver.

All refractive indices are solved on the same $128^3$ computational grid, which is sufficiently fine to satisfy the default discretization criterion over the complete refractive-index domain. Scattering quantities are evaluated using $180$ angular sampling points in a fixed scattering plane. The same numerical parameters are then used for the corresponding independent BiCG$_{\mathrm{CS}}$ calculations.

The complete shifted sweep can be launched from the shell as

\begin{lstlisting}[style=cli]
./adda -size 20 -grid 128 -pol fcd -int fcd -iter sbicg \
       -ntheta 180 -store_shifted_convergence -m $(
  for n in $(seq -f "%.3f" 1.300 0.025 1.800); do
    for k in $(seq -f "%.3f" 0.100 0.025 0.600); do
      printf "%s %s " "$n" "$k"
    done
  done
)
\end{lstlisting}

The implementation was evaluated using sequential CPU (SEQ), MPI, OpenCL (OCL), and the experimental OCL\_BLAS backend. The latter performs the Krylov linear algebra directly on the GPU, whereas in the standard OCL backend only the matrix--vector product is offloaded.

The performance measurements were carried out on an NVIDIA A100 GPU with \qty{80}{\GiB} of memory, paired with a dual-socket AMD EPYC 7543 node with \qty{480}{\GiB} of host memory for the GPU runs, and on dual-socket AMD EPYC 9654 Genoa nodes with \qty{768}{\GiB} of memory per node for the SEQ and MPI runs. While most MPI runs were performed on a single node, additional configurations with 128 MPI ranks distributed across two and four nodes were also investigated. To minimize variability caused by resource sharing, all simulations were performed using exclusive node allocation, ensuring that the requested CPU or GPU resources were not shared with other jobs. The software stack consisted of GCC~12.2, MPICH~3.4a2, and FFTW~3.3.10. \adda was compiled in release mode using the default \texttt{-Ofast} optimization flag.

%-------------------------------------------------------------------------------
\subsection{Timing and performance metrics}
%-------------------------------------------------------------------------------

The total execution time of an \adda calculation can be decomposed into three main phases: initialization, solver (named internal-fields in \adda), and scattering calculations (postprocesing),

\begin{equation}
    T_{\mathrm{tot}}
    =
    T_{\mathrm{init}}
    +
    T_{\mathrm{solv}}
    +
    T_{\mathrm{scat}}.
\end{equation}
Here, $T_{\mathrm{init}}$ includes the construction of the target, initialization of the interaction operator and FFT structures, and, for the OpenCL backends, initialization of the device (GPU) environment. In the shifted calculation, these operations are common to all refractive indices and are therefore performed only once. Consequently, this phase benefits from a speedup of approximately $N_m$.

The scattering stage, $T_{\mathrm{scat}}$, contains the evaluation of integrated quantities, such as cross sections, as well as angle-resolved quantities obtained from the scattered fields. The cost of the latter depends on the requested angular resolution and therefore on parameters such as \texttt{-ntheta}. Unlike the Krylov process, the scattering calculation is performed separately for each refractive index. Its computational cost therefore increases linearly with $N_m$, and no speedup is expected.

The solver stage, $T_{\mathrm{solv}}$, corresponds to the solution of the DDA linear systems, which is a computational bottleneck in most DDA simulations. To analyze the solver performance in more detail, the time is further decomposed as

\begin{equation}
    T_{\mathrm{solv}}
    =
    T_{\mathrm{MVP}}
    +
    T_{\mathrm{other}},
\end{equation}
where $T_{\mathrm{MVP}}$ is the time spent in matrix--vector products and $T_{\mathrm{other}}$ contains the remaining solver operations. For the OpenCL backends, the latter also includes the data transfers (to and from the device).

For the independent approach using BiCG$_{\mathrm{CS}}$, the timing associated with each component is summed over all refractive indices, whereas the corresponding shifted timing is obtained from a single simultaneous SBiCG$_{\mathrm{CS}}$ calculation. The speedup associated with a timing component $X$ is therefore defined as

\begin{equation}
    S_X
    =
    \frac{T_X^{\mathrm{BiCG}}}
         {T_X^{\mathrm{SBiCG}}}.
\end{equation}

We report the speedups associated with the MVPs, the remaining solver operations, the complete solver solution, and the total wall-clock time. The wall-clock speedup provides the most direct measure of the practical acceleration experienced by the user, but varies a lot with problem parameters, whereas the component-wise speedups identify which parts of the calculation benefit from the shifted formulation and which become limiting after acceleration.

Finally, to distinguish the purely algorithmic reduction from variations in the execution time of an individual MVP, we also report the MVP-count reduction defined in Eq.~\eqref{eq:mvp_count_speedup}.

%-------------------------------------------------------------------------------
\section{Results and discussion}
\label{sec:results}
%-------------------------------------------------------------------------------
\subsection{Numerical validation}
\label{sec:results_accuracy}
%-------------------------------------------------------------------------------

Before analyzing the computational performance, we verify that the shifted formulation reproduces the optical quantities obtained from independent BiCG$_{\mathrm{CS}}$ calculations. This numerical-equivalence analysis is performed in sequential mode, which is used here as a representative reference implementation. The validity of the shifted formulation was verified to be not specific to this backend (data not shown), since the same linear systems and solver algorithm are used with the other computational backends, although small backend-dependent differences may arise from floating-point operation ordering. 

\begin{figure}[ht]
    \centering
    \includegraphics[width=\linewidth]{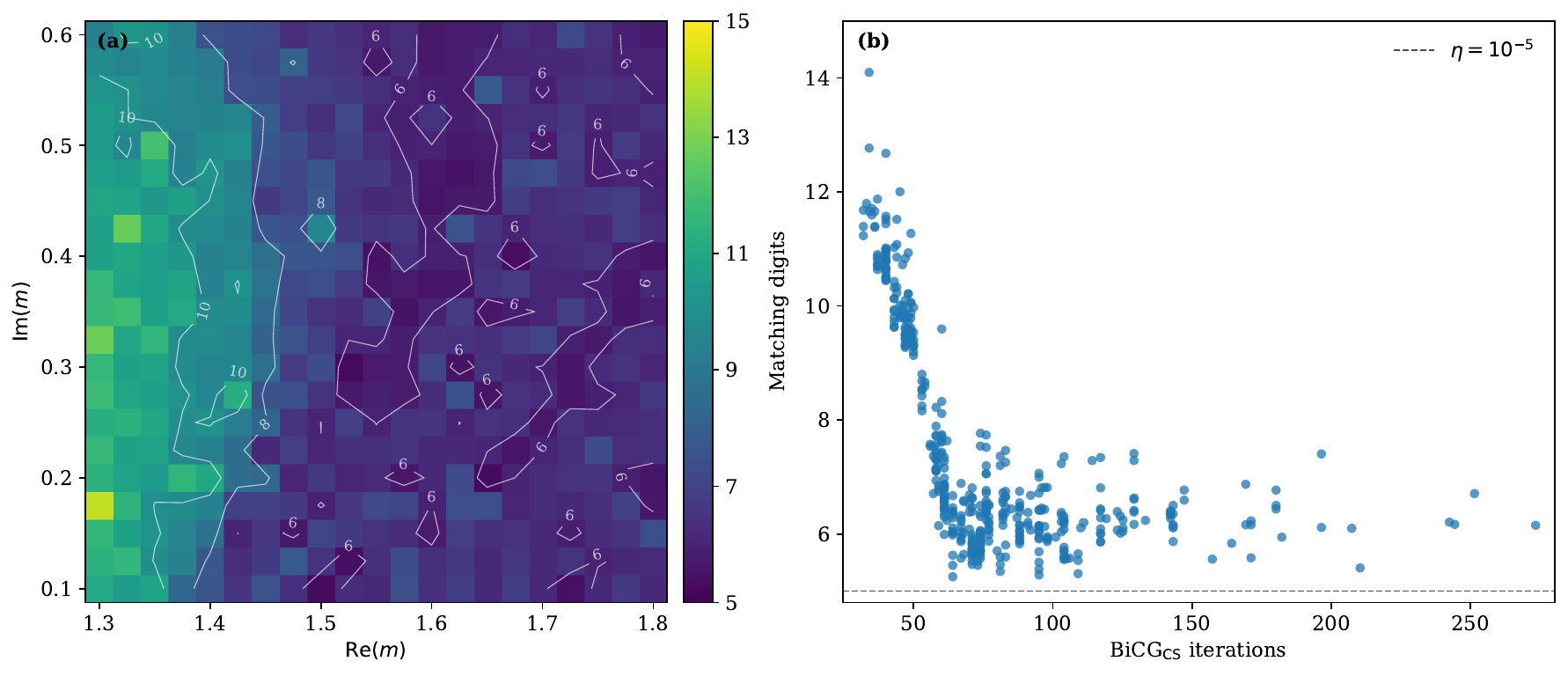}
    \caption{Numerical agreement between the shifted SBiCG$_{\mathrm{CS}}$ solver and independent BiCG$_{\mathrm{CS}}$ calculations in sequential mode for the $21\times21$ refractive-index grid. (a) Estimated number of matching digits in the extinction cross section over the complex plane. Larger values indicate closer agreement. (b) Same quantity versus the number of iterations required by the independent BiCG$_{\mathrm{CS}}$ calculation. The dashed horizontal line indicates the solver tolerance $\eta=\num{e-5}$.}
    \label{fig:cext_equivalence}
\end{figure}

Figure~\ref{fig:cext_equivalence} shows the relative difference in the extinction cross section, defined by Eq.~\eqref{eq:cext_error}, over the complete refractive-index domain. The agreement ranges from machine precision for the most favorable cases to differences of order \num{e-6} for the largest discrepancies. The maximum relative difference is \num{5.5e-6}, still below the prescribed solver tolerance $\eta=\num{e-5}$ over the complete investigated domain. 

Differences close to machine precision are observed for low-index regions, whereas larger discrepancies generally appear toward larger $\Repart m$. This behavior is related to the length of the Krylov sequence, as confirmed by Fig.~\ref{fig:cext_equivalence}(b), where the agreement generally decreases as the number of iterations increases. This trend is consistent with the finite-precision effects discussed in Section~\ref{sec:numerical_equivalence}. Although such effects have been studied for Krylov methods, particularly regarding the gap between recursively computed and true residuals~\cite{GUTKNECHT2000}, their implications for the agreement between different solvers are less clear. The precise mechanism responsible for the discrepancies observed here therefore remains to be established.

As an additional verification, the calculation was repeated with the much stricter tolerance $\eta=\num{e-16}$, for which the relative differences in $C_{\mathrm{ext}}$ remain within \num{e-14} (data not shown). Overall, the above results confirm the correct implementation of the shifted solver.

%-------------------------------------------------------------------------------
\subsection{Scaling and saturation with the number of shifted systems}
\label{sec:results_scaling}
%-------------------------------------------------------------------------------

In this section, all shifted systems correspond to the same refractive index and therefore have identical convergence histories. This configuration isolates the intrinsic scaling of the shifted algorithm from variations in the number of iterations across the refractive-index domain.

\begin{figure}[ht]
    \centering
    \includegraphics[width=\linewidth]{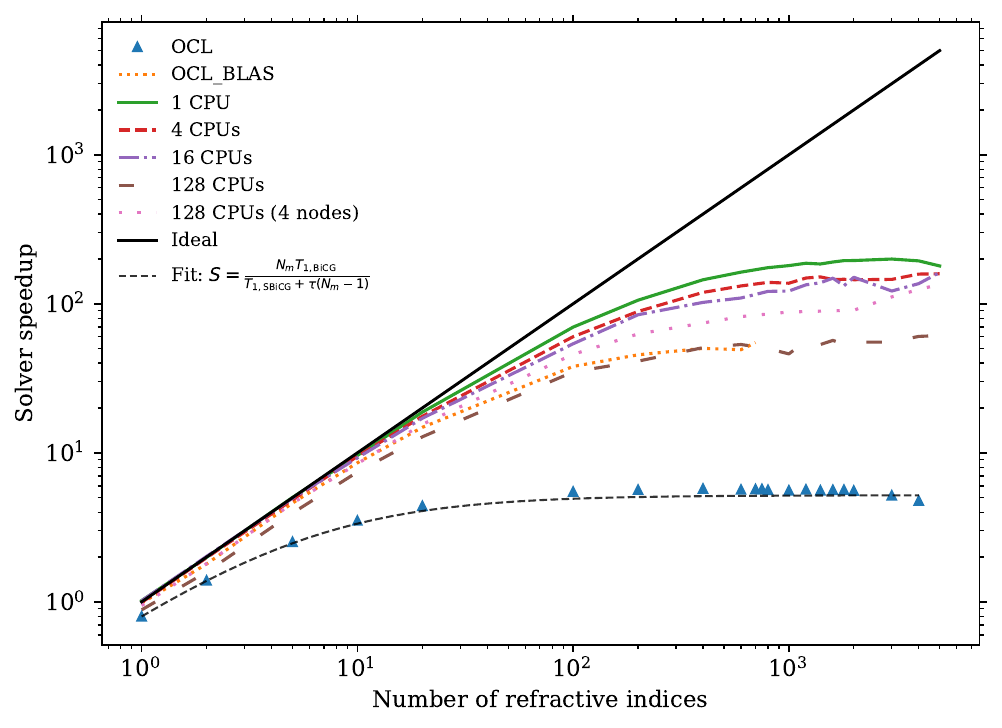}
    \caption{Solver speedup of the shifted SBiCG$_{\mathrm{CS}}$ solver as a function of the number of simultaneously treated systems, $N_m$, for the different computational backends. The physical and numerical parameters are the same as those defined in Section~\ref{sec:benchmark_configuration}. All shifted systems are identical and use the refractive index $m=1.8+0.6\mathrm{i}$. The dashed black line denotes the ideal scaling, $S=N_m$. The OCL fit illustrates the model of Eq.~\eqref{eq:shifted_speedup_model}.}
    \label{fig:scaling}
\end{figure}

For small $N_m$, Fig.~\ref{fig:scaling} shows that the solver speedup remains close to the ideal behavior, $S=N_m$. In this regime, the computational cost is dominated by the shared MVPs, and the shifted calculation therefore approaches the ideal reduction predicted by Eq.~\eqref{eq:ideal_speedup}. As $N_m$ increases, however, the cost of the shift-dependent vector and scalar operations becomes progressively more important, causing the speedup to depart from ideal linear scaling and eventually saturate.

The solver time of the shifted calculation can be described by

\begin{equation}
    T_{\mathrm{SBiCG}}(N_m)
    =
    T_{\mathrm{s},1}+\tau(N_m-1),
    \label{eq:shifted_time_model}
\end{equation}
where $T_{\mathrm{s},1}$ is the measured solver time of SBiCG$_{\mathrm{CS}}$ for a single system ($N_m=1$), and $\tau$ represents the incremental cost associated with each additional shifted system. Note that the results for two largest values of $N_m$ for OCL and the largest $N_m$ for SEQ in Fig.~\ref{fig:scaling} indicate the need for an extra term, scaling faster than linear, to be added to \eqref{eq:shifted_time_model} when the occupied memory approaches the maximal capacity, but we do not investigate it further.

For comparison with independent calculations, we define $T_1$ as the measured solver time of a single system solved using the BiCG$_{\mathrm{CS}}$ solver. A direct comparison between $T_1$ and $T_{\mathrm{s},1}$ shows differences below $2\%$ for SEQ, MPI, and OCL\_BLAS, whereas the differences reach $13\%$ for MPI with 128 ranks on a single node and $25\%$ for OCL. The larger discrepancy observed for OCL arises because only the MVP is accelerated on the GPU, whereas the additional vector operations required by SBiCG$_{\mathrm{CS}}$ are performed on the CPU, as discussed in Section~\ref{sec:shifted_lanczos}. However, one-third of this discrepancy can be attributed to an unnecessary vector operations which has been removed during futher code development.

For each backend, $T_{\mathrm{s},1}$ is fixed to its measured value at $N_m=1$, while $\tau$ is determined by a least-squares fit of Eq.~\eqref{eq:shifted_time_model} over all measured values of $N_m$. The obtained parameters are summarized in Table~\ref{tab:scaling_fit}, with $R^2>0.95$ in all cases.

Assuming comparable execution times for the individual systems, $N_m$ independent BiCG$_{\mathrm{CS}}$ calculations require approximately $N_mT_1$. The corresponding speedup predicted by Eq.~\eqref{eq:shifted_time_model} is therefore

\begin{equation}
    S(N_m)
    =
    \frac{N_mT_1}
         {T_{\mathrm{s},1}+\tau(N_m-1)},
    \label{eq:shifted_speedup_model}
\end{equation}

The distinction between $T_1$ and $T_{\mathrm{s},1}$ accounts for the lower speedup observed for OCL at small $N_m$, where the additional overhead of SBiCG$_{\mathrm{CS}}$ is more significant. As $N_m$ increases, the contribution of this initial overhead becomes progressively less important, yielding the asymptotic solver speedup

\begin{equation}
    S_{\mathrm{sat}}
    =
    \lim_{N_m\rightarrow\infty}S(N_m)
    =
    \frac{T_1}{\tau}.
\end{equation}

\begin{table}[ht]
\centering
\caption{Parameters obtained by fitting the shifted solver execution time with Eq.~\eqref{eq:shifted_time_model}. $T_{\mathrm{s},1}$ is fixed to the measured SBiCG$_{\mathrm{CS}}$ single-system time, while $\tau$ is the fitted incremental cost per additional shifted system. $T_1$ is the measured independent BiCG$_{\mathrm{CS}}$ single-system time, used to compute the asymptotic speedup $S_{\mathrm{sat}}=T_1/\tau$.}
\label{tab:scaling_fit}
\begin{tabular}{lrrrr}
\toprule
Backend & $T_1$ (s) & $T_{\mathrm{s},1}$ (s)
        & $\tau$ (s) & $S_{\mathrm{sat}}$ \\
\midrule
SEQ               & 133  & 130  & 0.66  & 201 \\
MPI 4             & 34   & 35   & 0.21  & 162 \\
MPI 16            & 9.4  & 9.3  & 0.063 & 150 \\
MPI 128 (1 node)  & 2.1  & 2.4  & 0.035 & 60  \\
MPI 128 (2 nodes) & 1.8  & 2.0  & 0.028 & 65  \\
MPI 128 (4 nodes) & 1.8  & 2.0  & 0.015 & 126 \\
OCL               & 4.4  & 5.5  & 0.85  & 5.2 \\
OCL\_BLAS         & 1.9  & 2.0  & 0.034 & 57  \\
\bottomrule
\end{tabular}
\end{table}

The asymptotic speedup decreases as the common computation is increasingly accelerated, from $S_{\mathrm{sat}}=201$ in SEQ to 60 with 128 MPI ranks on a single node, reflecting the increasing relative contribution of shift-dependent operations. Distributing the same 128 MPI ranks across multiple nodes reduces the incremental cost from $\tau=\qty{35}{\ms}$ on a single node to $\qty{28}{\ms}$ and $\qty{15}{\ms}$ on two and four nodes, respectively. Meanwhile, the corresponding asymptotic speedups increase to 65 and 126. This suggests that memory contention for shared hardware resources plays an important role, i.e., the memory bandwidth of a single node becomes a bottleneck. By contrast, increasing the computational grid to $256^3$ does not significantly affect the speedup on a single node (data not shown), suggesting that latency of MPI operations is negligible. Still, latency may become important for smaller grids.

The effect of the increase of shift-dependent operations is even more pronounced for OCL, where they are neither executed on the GPU nor parallelized, limiting the asymptotic speedup to $S_{\mathrm{sat}}=5.2$. OCL\_BLAS reduces this limitation by performing the linear algebra directly on the GPU. The incremental cost decreases to only $\tau=\qty{34}{\ms}$ per additional shifted system, increasing the asymptotic speedup to 57. This value nevertheless remains below those obtained with the sequential and moderate-MPI configurations since the MVP is also greatly accelerated. The fact that MVP (i.e., FFT) is accelerated by the GPU more than the remaining linear-algebra part is probably specific to the used device and requires further investigation. 

These results also highlight a trade-off between computational performance and memory usage. In the OCL backend, only the MVP is offloaded to the GPU, while the two shift-dependent solver vectors remain in host (CPU) memory (see~\ref{app:memory_model}). In OCL\_BLAS, by contrast, these vectors are stored on the GPU. Consequently, in the OCL backend, the GPU-memory requirement is independent of $N_m$, while the practical limit of $N_m$ is determined by the available CPU-memory (as for SEQ and MPI backends). Whereas for OCL\_BLAS this limit is set by the available GPU memory, which is typically smaller. For example, in the configuration considered here, this limit is reached slightly above $N_m=700$, corresponding to \qty{70}{\GiB} of total device-memory usage and about \qty{34}{\GiB} for the largest individual buffer.

The maximum size of a single OpenCL allocation must also be considered, if it is smaller than half of the available device memory, since the shift-dependent storage is distributed over two equally sized arrays. Unfortunately, the effective maximum allocation size is hardware- and implementation-dependent. Usually, only a quarter of device memory is guaranteed (the minimum mandated by the OpenCL standard), while larger allocations may still succeed in practice. This was the case on our NVIDIA A100 with allocations of up to a half of device memory. On devices with a lower single-allocation limit, the shift-dependent arrays can be split into several smaller buffers.

Memory is generally less restrictive for the SEQ and MPI backends. In particular, the MPI implementation, which is unique to \adda among the major open-source DDA codes~\cite{ARGENTIN2026}, combines the memory across several compute nodes, reaching values much larger than that of a single GPU. For large $N_m$, one should thus compare a single MPI run (for the whole batch) versus several OCL\_BLAS runs (for smaller sub-batches). Note, however, that even if having larger wall-times OCL\_BLAS backend is expected to consume less electrical energy~\cite{ARGENTIN2026}.

%-------------------------------------------------------------------------------
\subsection{Performance over the refractive-index domain}
%-------------------------------------------------------------------------------

The previous experiment considered shifted systems with identical convergence histories in order to isolate the intrinsic scaling of the algorithm. We now return to the $21\times21$ refractive-index sweep used for the numerical-equivalence analysis in Section~\ref{sec:results_accuracy}, for which the number of iterations varies substantially across the refractive-index domain. The results are summarized in Table~\ref{tab:performance}.

In sequential mode, the independent calculations require a total of \num{34263} MVPs, with individual iteration counts ranging from 32 to 273 and a mean of 78, yielding an ideal speedup of $S_{\mathrm{ideal}}=126$ from Eq.~\eqref{eq:ideal_speedup}. The corresponding shifted systems converge between 32 and 256 iterations, requiring 256 shared MVPs and yielding an observed MVP-count reduction of $S_{\mathrm{count}}=134$ from Eq.~\eqref{eq:mvp_count_speedup}. The difference between the 256 and 273 iterations arise from finite-precision effects, as discussed in Section~\ref{sec:shifted_lanczos}. Small backend-dependent variations in floating-point arithmetic similarly lead to corresponding $S_{\mathrm{count}}$ values ranging from 126 to 134. These results confirm that the dominant matrix--vector operation is effectively shared across all refractive indices.

The reduction in MVP cost does not translate directly into the same acceleration of the complete solver because the shift-dependent vector and scalar operations remain. In sequential mode, the solver is accelerated by a factor of 83. The corresponding speedups decrease to 74, 68, and 37 for 4, 16, and 128 MPI ranks, respectively. This behavior is consistent with the scaling analysis of Section~\ref{sec:results_scaling}.

The difference between the two GPU implementations is even more pronounced. OCL provides a solver speedup of only 4.7 despite a MVP speedup of 116, while OCL\_BLAS increases the solver speedup back to impressive level of 38.

\begin{table*}[ht]
\centering
\caption{Performance of SBiCG$_{\mathrm{CS}}$ for the $21\times21$ refractive-index sweep. $S_{\mathrm{count}}$ denotes the observed reduction in the number of matrix--vector products, defined by Eq.~\eqref{eq:mvp_count_speedup}. For each timing component, $T$ is the execution time of the complete shifted calculation and $S$ is the corresponding speedup relative to the sum of the 441 independent BiCG$_{\mathrm{CS}}$ calculations.}
\label{tab:performance}
\resizebox{\textwidth}{!}{%
\begin{tabular}{lr|rr|rrrrrr|rr|rr}
\toprule
& &
\multicolumn{2}{c|}{Initialization} &
\multicolumn{6}{c|}{Solver} &
\multicolumn{2}{c|}{Scattering} &
\multicolumn{2}{c}{Wall time} \\

\cmidrule(lr){3-4}
\cmidrule(lr){5-10}
\cmidrule(lr){11-12}
\cmidrule(lr){13-14}

& & & &
\multicolumn{2}{c}{MVP} &
\multicolumn{2}{c}{Other} &
\multicolumn{2}{c|}{Total} &
& & & \\

\cmidrule(lr){5-6}
\cmidrule(lr){7-8}
\cmidrule(lr){9-10}

Backend
&
$S_{\mathrm{count}}$ &
$T$ (s) &
$S$ &
$T$ (s) &
$S$ &
$T$ (s) &
$S$ &
$T$ (s) &
$S$ &
$T$ (s) &
$S$ &
$T$ (s) &
$S$ \\
\midrule

SEQ
& 134
& 5.0 & 440
& 347 & 135
& 224 & 2.3
& 571 & 83
& 719 & 1.0
& \num{1311} & 39 \\

MPI 4
& 134
& 1.7 & 400
& 92 & 131
& 75 & 3.4
& 167 & 74
& 249 & 1.0
& 423 & 31 \\

MPI 16
& 134
& 0.9 & 354
& 25 & 131
& 24 & 3.0
& 48 & 68
& 67 & 1.0
& 120 & 31 \\

MPI 128
& 133
& 0.8 & 335
& 6.3 & 118
& 15 & 1.7
& 21 & 37
& 8.1 & 1.2
& 32 & 35 \\

OCL
& 126
& 2.8 & 430
& 8.8 & 116
& 333 & 1.8
& 342 & 4.7
& 583 & 1.0
& 941 & 3.5 \\

OCL\_BLAS
& 128
& 2.8 & 433
& 4.6 & 130
& 15 & 9.3
& 19 & 38
& 575 & 1.0
& 606 & 4.2 \\

\bottomrule
\end{tabular}%
}
\end{table*}

%-------------------------------------------------------------------------------
\subsection{Computational bottlenecks and practical considerations}
%-------------------------------------------------------------------------------

The component-wise speedups in Table~\ref{tab:performance} illustrate an important consequence of accelerating the iterative solution: the computational bottleneck progressively shifts toward stages that are not accelerated by the shifted Krylov method. The initialization phase is shared, with speedups of several hundred, while the scattering phase is not. As a result, for the CPU configurations, the complete wall-clock speedup remains between 31 and 39, considerably below the MVP speedup of more than 100.

This bottleneck shift is even more pronounced for the GPU backends. For OCL, the MVP represents less than 1\% of the shifted wall time, whereas the non-MVP solver operations account for 35\% and the scattering calculation for 62\%. OCL\_BLAS largely removes the solver-side limitation. As a consequence, however, the scattering calculation becomes overwhelmingly dominant, representing 95\% of the complete shifted execution time. This explains why the solver speedup of 38 obtained with OCL\_BLAS translates into a wall-clock speedup of only 4.2 for the complete calculation.

These results emphasize that the performance of the shifted solver should not be assessed from the MVP speedup alone. Once the linear-system solution has been accelerated, further reductions of the total execution time require optimization of the remaining stages. In particular, GPU acceleration of the scattered-field evaluation and the projection strategy described in~\ref{app:scattering} provide a natural route toward recovering a larger fraction of the solver-level acceleration at the scale of the complete DDA simulation.

The present benchmark was deliberately chosen as a demanding end-to-end test. It combines a broad refractive-index range, producing a large variation in convergence rate, with 180 scattering angles, which makes the non-accelerated scattering stage significant. The resulting wall-clock speedups should therefore be interpreted as realistic but sample values -- both larger and smaller values are possible depending on the problem parameters.

In particular, both computational time and memory requirements (see \ref{app:memory_model}) for each extra shift scale with $N$, while that for shared MVP - with $N_{\mathrm{box}}$. Thus, the shifted method will be even more efficient in terms of speedup and maximum $N_m$ achievable on a given hardware for porous particles, i.e., when $N\ll N_{\mathrm{box}}$.

From an algorithmic point of view, the most efficient strategy is to treat as many refractive indices as possible in a single shifted calculation, since the shared Krylov process is then governed only by the slowest-converging system. Splitting the sweep into several batches necessarily repeats part of this work and therefore reduces the attainable speedup. Whenever memory permits (which is often the case for SEQ and, especially, MPI backends), a single large batch should thus be preferred.

For OCL\_BLAS, the available GPU memory may impose a maximum batch size. In this case, it is recommended to group systems with similar expected iteration counts to reduce the sum of the batch maxima. An approximate convergence map can therefore be useful when planning large GPU sweeps. Such information may be obtained from inexpensive preliminary calculations using a coarser discretization or from previous iterative-convergence results for similar particles. For sufficiently small particles ($x_p\leq1$), analytical estimates such as those discussed in Ref.~\cite{YURKIN2013} may be used.

%-------------------------------------------------------------------------------
\section{Conclusion}
\label{sec:conclusion}
%-------------------------------------------------------------------------------

We implemented in \adda a shifted complex-symmetric Krylov solver for accelerating refractive-index sweeps in DDA. For homogeneous and isotropic particles with fixed discretization and illumination, the refractive-index dependence can be expressed as a scalar shift of the interaction matrix, allowing all systems to share a single Krylov process. The resulting SBiCG$_{\mathrm{CS}}$ implementation reproduces independent BiCG$_{\mathrm{CS}}$ calculations within the prescribed iterative-method tolerance, while reducing the number of MVPs by a factor of 130 for the broad sweep of 441 refractive-index values considered here. End-to-end simulation time was accelerated by up to 40 times, but this value depends on several problem parameters. The ability to perform such sweep in a single run is tightly related to the simple structure of the DDA linear system (or volume integral equation in general) and its dependence on the refractive index, which explains why no similar capabilities have been reported for any other light-scattering methods.

From a practical perspective, the method is particularly attractive whenever many refractive indices must be evaluated for the same particle configuration, for example for optical-property databases, or sensitivity studies. Users should treat as many refractive indices as possible in a single shifted calculation, since splitting the sweep into several batches repeats part of the shared Krylov work. And even for a batch of a few refractive indices the shifted solver provides the corresponding few-times acceleration essentially for free. Moreover, the acceleration will be even better for porous or aggregated particles.

Results show that once the solver solution is strongly accelerated, the computational bottleneck moves to stages that are not shared by the shifted formulation. In the present implementation, the scattered-field calculation can dominate the complete execution time, especially for GPU backends and calculations requiring many scattering directions. GPU acceleration and projection-based reduction of this phase (with already available prototypes) constitute the next step toward increasing the wall-clock speedup.

Another route for improvement is to combine shifted Krylov methods with reusable particle representations such as T-matrix methods or multiple-right-hand-side strategies, which could extend the acceleration to broader classes of related DDA simulations. For the latter, a prototype Block-Krylov implementation is already available at \url{https://github.com/inzhevatkin/adda/tree/Bicg_block}, providing a possible basis for future combination with the shifted approach. On GPU backends, the memory limitation could also be mitigated by using single precision and by further reducing the memory footprint of the Green tensor through exploitation of the additional symmetry along the $x$-axis. Prototype implementations of these memory-reduction strategies are already available at \url{https://github.com/michelgross34/adda-cuda}, providing a basis for their future integration.

%%----------------------------------------------------------------------------%%
\section*{Data and code availability}
%-------------------------------------------------------------------------------

The exact version of \adda used in this study is available at \url{https://github.com/Argencle/adda/tree/35deda91}. The corresponding development branch will be merged into the main \adda repository available at \url{https://github.com/adda-team/adda}.

All data and scripts used to produce the results and figures in this study are publicly available at Zenodo: \url{https://doi.org/10.5281/zenodo.23983951}.

%%----------------------------------------------------------------------------%%
\section*{CRediT authorship contribution statement}
%-------------------------------------------------------------------------------

\textbf{Cl\'ement Argentin}: Conceptualization, Software, Validation, Investigation, Data curation, Writing -- original draft, Visualization. 
\textbf{Konstantin G. Inzhevatkin}: Conceptualization, Software, Validation, Investigation. 
\textbf{Maxim A. Yurkin}: Conceptualization, Software, Validation, Writing -- review \& editing, Supervision, Funding acquisition.

%-------------------------------------------------------------------------------
\section*{Acknowledgments}
%-------------------------------------------------------------------------------

This work was supported by the Normandy Region (project RADDAERO). Part of the simulations were performed using computing resources of CRIANN (Normandy, France).

%-------------------------------------------------------------------------------
\appendix
%-------------------------------------------------------------------------------
\section{Memory model}
\label{app:memory_model}
%-------------------------------------------------------------------------------

The memory overhead of the shifted solver can be estimated from the standard \adda memory model. For a conventional BiCG$_{\mathrm{CS}}$ calculation in free space,  the total memory requirement is~\cite{YURKIN2011} 
\begin{equation}
    M_{\mathrm{BiCG}}
    =
    (A N_\mathrm{box}
    +
    271 N)\unit{\byte},
    \label{eq:memory_bicg}
\end{equation}
where

\begin{equation}
    A
    =
    288
    +
    384\frac{n_p}{n_x}
    +
    \left(\frac{192}{n_p}\right)_{\mathrm{parallel}},
    \label{eq:A_CPU}
\end{equation}
and $n_p$ denotes the number of processors. The coefficient $A$ describes the memory required to perform MVPs, where the last term is present only for parallel MPI calculations. The coefficient 271 in Eq.~\eqref{eq:memory_bicg}  follows from the general expression $31+48(n_{\mathrm{vec}}+1)$ with $n_{\mathrm{vec}}=4$ for BiCG$_{\mathrm{CS}}$.

Defining the target occupancy as

\begin{equation}
    \phi
    =
    \frac{N}{N_{\mathrm{box}}},
\end{equation}
i.e., $\phi=1$ for a cubic target and $\phi=\pi/6$ for a sphere in the computational box, Eq.~\eqref{eq:memory_bicg} can be rewritten as

\begin{equation}
    M_{\mathrm{BiCG}}
    =
    \left(
        \frac{A}{\phi}
        +
        271
    \right)
    N\unit{\byte}.
    \label{eq:M_bicgcs}
\end{equation}

The additional memory $\Delta M$ relative to a conventional BiCG$_{\mathrm{CS}}$ is given in Eq.~\eqref{eq:delta_M_CPU}, thus the relative memory overhead is

\begin{equation}
    \frac{\Delta M}
         {M_{\mathrm{BiCG}}}
    =
    \frac{
        96(N_m+1.5)
    }{
        A/\phi+271
    }.
\end{equation}
This expression corresponds to the optimized implementation, which requires three additional common vectors relative to BiCG$_{\mathrm{CS}}$. The previous version, used in benchmarks, had 1.5 replaced by 2.

Introducing the dimensionless coefficient

\begin{equation}
    C(\phi,n_x,n_p)
    =
    \frac{
        A/\phi+271
    }{96},
    \label{eq:C_main}
\end{equation}
the total memory requirement of the shifted solver can be expressed in the compact form

\begin{equation}
    \frac{M_{\mathrm{SBiCG}}}
         {M_{\mathrm{BiCG}}}
    =
    1+
    \frac{N_m+1.5}{C}.
\end{equation}
Thus, for the $128^3$ spherical target considered in the present study, the resulting values of $C$ are 

\begin{equation}
    C
   =
    \begin{cases}
        8.61,  & \mathrm{sequential\ mode}, \\
        12.4, & n_p=1,\\
        9.75,  & n_p=4,\\
        9.75,  & n_p=16,\\
        16.2, & n_p=128,
    \end{cases}
\end{equation}
with variation of less than a factor of two. In sequential mode, the dependence on the grid size is comparatively weak. For a spherical target, the coefficient approaches an asymptotic value of $8.55$ as the grid size increases. An obvious consequence of this model is that sparsely occupied targets with low $\phi$ have larger value of $C$ and, hence, can use larger $N_m$ for a given memory overhead.

The memory organization of the OpenCL backends is different, since two distinct types of memory must now be considered, namely the host memory $M_{\mathrm{CPU}}$ and the device memory $M_{\mathrm{GPU}}$. Moreover, both these parts contain the memory allocated by the OpenCL framework, which is not directly controlled by \adda and, hence, cannot be calculated from first principles.

With OCL backend, the device-memory requirement is independent of $N_m$, since only the MVP (the first term in Eq.~\eqref{eq:memory_bicg}) is offloaded to the GPU. The increase relative to BiCG$_{\mathrm{CS}}$ is therefore contained in the CPU memory and follows the same Eq.~\eqref{eq:delta_M_CPU}. However, the corresponding coefficient $C_{\mathrm{OCL}}$ differs from that of the sequential backend because the total CPU memory $M_{\mathrm{CPU}}^{\mathrm{BiCG}}$ differs as well. Still, those two are connected analogous to Eqs.~\eqref{eq:M_bicgcs}--\eqref{eq:C_main}:

\begin{equation}
    C_{\mathrm{OCL}}
    =
    \frac{
        M_{\mathrm{CPU}}^{\mathrm{BiCG}}
    }{
        96N
    },
\end{equation}

\begin{equation}
    \frac{
        M_{\mathrm{CPU}}^{\mathrm{SBiCG}}
    }{
        M_{\mathrm{CPU}}^{\mathrm{BiCG}}
    }
    =
    1+
    \frac{N_m+1.5}{C_{\mathrm{OCL}}}.
    \label{eq:memory_ratio_OCL}
\end{equation}

The measured host-memory requirement can be described accurately by adding a grid-independent contribution $M_0^\mathrm{CPU}$ (due to the OpenCL framework) to the remaining part of the analytical \adda memory model in Eq.~\ref{eq:memory_bicg},

\begin{equation}
    M_{\mathrm{CPU}}^{\mathrm{BiCG}}
    =
    M_0^\mathrm{CPU}+271N\unit{\byte},
\end{equation}
giving the corresponding coefficient

\begin{equation}
    C_{\mathrm{OCL}}
    =
    \frac{271}{96}
    +
    \frac{M_0^\mathrm{CPU}}{96N}.
    \label{eq:C_OCL}
\end{equation}

For the $128^3$ grid, $M_0^\mathrm{CPU}=130$~MiB (on the used hardware), and $C_{\mathrm{OCL}}=4.12$. As the discretization size increases, this fixed contribution becomes progressively less important relative to the $\mathcal{O}(N)$ solver storage. Consequently,

\begin{equation}
    C_{\mathrm{OCL}}^{\infty}
    =
    \frac{271}{96}
    =
    2.82.
\end{equation}
Note that this asymptotic value and Eq.~\eqref{eq:C_OCL} are independent of the particle shape, since the memory is expressed solely in terms of the number of nonvoid voxels $N$.

The device-memory requirement of the OpenCL backends can be determined from the buffers allocated by \adda. The MVP-related allocations consist of the field array, the Fourier-transformed interaction tensor, and two FFT working arrays. For a cubic computational grid of size $n_x^3$, with FFT dimensions $G_x=G_y=G_z=2n_x$, this gives

\begin{equation}
    M_{\mathrm{MVP}}
    =
    A_{\mathrm{GPU}}(n_x,b)N_{\mathrm{box}}\unit{\byte},
\end{equation}
with

\begin{equation}
    A_{\mathrm{GPU}}(n_x,b)
    =
    96
    +
    192\left(1+\frac{1}{n_x}\right)^2
    +
    \frac{384b}{n_x},
    \label{eq:A_GPU}
\end{equation}
where $b$ denotes the number of slices processed simultaneously during the FFT operations. For the default configurations considered here, $b=32$, whereas the \adda memory-saving option \texttt{-opt mem} reduces this number to $b=1$.

In addition to these MVP-related buffers, the OCL implementation allocates two complex vectors of $48N$ bytes each, the dipole-position array of $6N$ bytes, and the material-index array of $N$ bytes. The corresponding occupied-voxel-dependent contribution is therefore $103N\unit{\byte}$. Thus, the device-memory requirement of a calculation with OCL is

\begin{equation}
    M_{\mathrm{GPU}}^{\mathrm{OCL}}
    =
    \left[
        A_{\mathrm{GPU}}N_{\mathrm{box}}
        +
        103N
    \right]\unit{\byte}
    +M^\mathrm{GPU}_\mathrm{0},
\end{equation}
where $M^\mathrm{GPU}_\mathrm{0}$ denotes the additional memory used by the OpenCL runtime and GPU driver and is not tracked by \adda (similar to $M^\mathrm{CPU}_\mathrm{0}$ above). This contribution depends on the hardware and software environment and is usually around 100~MiB.

In OCL\_BLAS, the shift-dependent linear-algebra operations are instead performed directly on the GPU. The two arrays containing $\mathbf{x}^{(s)}_k$ and $\mathbf{p}^{(s)}_k$ are therefore stored in device memory. In the optimized implementation, the additional GPU memory relative to BiCG$_{\mathrm{CS}}$ is

\begin{equation}
    \Delta M_{\mathrm{GPU}}^{\mathrm{OCL\_BLAS}}
    =
    96(N_m-1)N\unit{\byte},
    \label{eq:delta_M_OCL_BLAS}
\end{equation}
instead of Eq.~\eqref{eq:delta_M_CPU}. The first shifted system reuses storage already present, while each subsequent system requires two additional full-size complex vectors. For comparison, the implementation used for the benchmarks required two additional temporary GPU vectors and, hence, lacking $-1$ in the above formula.

A device-memory coefficient can then be introduced as

\begin{equation}
    C_{\mathrm{OCL\_BLAS}}
    =
    \frac{
        M_{\mathrm{GPU}}^{\mathrm{BiCG}}
    }{
        96N
    },
    \label{eq:C_OCL_BLAS}
\end{equation}
such that

\begin{equation}
    \frac{
        M_{\mathrm{GPU}}^{\mathrm{SBiCG}}
    }{
        M_{\mathrm{GPU}}^{\mathrm{BiCG}}
    }
    =
    1+
    \frac{N_m-1}{C_{\mathrm{OCL\_BLAS}}}.
    \label{eq:memory_ratio_OCL_BLAS}
\end{equation}
which for the $128^3$ spherical target, gives the measured value of $C_{\mathrm{OCL\_BLAS}}=10.8$. As for OCL, this coefficient is grid-dependent.

To generalize this behavior to arbitrary target shapes and grids, the baseline device-memory requirement can be determined from the FFT-related buffers and the additional solver storage. Compared with OCL, conventional BiCG$_{\mathrm{CS}}$ in OCL\_BLAS requires four additional full-size complex-vector equivalents, including the clBLAS workspace. The occupied-voxel-dependent contribution is therefore $[103N+192N]\unit{\byte}=295N\unit{\byte}$, giving

\begin{equation}
    M_{\mathrm{GPU}}^{\mathrm{BiCG}}
    =
    \left[
        A_{\mathrm{GPU}}(n_x,b)N_{\mathrm{box}}
        +
        295N
    \right]\unit{\byte}
    +M^\mathrm{GPU}_\mathrm{0},
    \label{eq:GPU_memory_general}
\end{equation}
where $A_{\mathrm{GPU}}(n_x,b)$ is given by Eq.~\eqref{eq:A_GPU}. Equation~\eqref{eq:C_OCL_BLAS} therefore becomes

\begin{equation}
    C_{\mathrm{OCL\_BLAS}}
    =
    \frac{295}{96}
    +
    \frac{A_{\mathrm{GPU}}(n_x,b)}{96\phi}
    +\frac{M^\mathrm{GPU}_\mathrm{0}}{96N}.
\end{equation}

For fixed $b$ and $n_x\to\infty$, Eq.~\eqref{eq:A_GPU} gives $A_{\mathrm{GPU}}\to288$. The asymptotic device-memory coefficient is therefore

\begin{equation}
    C_{\mathrm{OCL\_BLAS}}^{\infty}(\phi)
    =
    \frac{295}{96}
    +
    \frac{288}{96\phi}
    =
    \frac{295}{96}
    +
    \frac{3}{\phi},
    \label{eq:C_OCL_BLAS_asymptotic}
\end{equation}
For a spherical target $C_{\mathrm{OCL\_BLAS}}^{\infty}=8.8$, slightly larger than the sequential asymptotic coefficient of 8.55 due to the additional solver storage on the GPU.

%-------------------------------------------------------------------------------
\section{Acceleration of the scattered-field calculation}
\label{app:scattering}
%-------------------------------------------------------------------------------

For a set of voxels with polarizations $\mathbf{p}_j$ located at $\mathbf{r}_j$, the far-field scattered electric field in a direction $\hat{\mathbf{n}}$ (where the hat denotes unit vectors) contains a sum of the form

\begin{equation}
    \mathbf{F}(\hat{\mathbf{n}})
    =
    \sum_{j=1}^{N}
    \mathbf{p}_j
    \exp\left(
        -\mathrm{i}k
        \hat{\mathbf{n}}\cdot\mathbf{r}_j
    \right).
    \label{eq:far_field_sum}
\end{equation}
A direct evaluation therefore requires a contribution from every voxel for every scattering direction.

For scattering directions restricted to a plane aligned with the computational grid, however, one coordinate of $\hat{\mathbf{n}}$ vanishes. Consider, for example, the $yz$ scattering plane, for which

\begin{equation}
    \hat{n}_x=0.
\end{equation}
For a regular lattice with voxel positions $\mathbf{r}_{lmn}=(x_l,y_m,z_n)$ (integer indices $l,m,n$ should not be confused with the same symbols in other parts of the paper), Eq.~\eqref{eq:far_field_sum} can then be rewritten as

\begin{align}
    \mathbf{F}(\hat{\mathbf{n}})
    &=
    \sum_{l,m,n}
    \mathbf{p}_{lmn}
    \exp\left[
        -\mathrm{i}k
        \left(
            \hat{n}_y y_m+
            \hat{n}_z z_n
        \right)
    \right]
    \nonumber\\
    &=
    \sum_{m,n}
    \left(
        \sum_l \mathbf{p}_{lmn}
    \right)
    \exp\left[
        -\mathrm{i}k
        \left(
            \hat{n}_y y_m+
            \hat{n}_z z_n
        \right)
    \right].
\end{align}

Thus, the voxel polarizations can first be projected along the direction normal to the scattering plane, after which the angular dependence is evaluated using only the resulting two-dimensional array. The same procedure applies to the $xz$ and $xy$ planes by summing first along the $y$ and $z$ directions, respectively. Hence, for a cubic grid of linear dimension $L$, the directional calculation is reduced from $\mathcal{O}(L^3)$ to $\mathcal{O}(L^2)$.

A prototype implementation of this approach, including the scattered-field calculation on the GPU, is available at \url{https://github.com/Argencle/adda/tree/feature/scattering}. This optimization is intended for future integration into the main \adda code as part of the ongoing development of GPU-accelerated scattered-field calculations.

While this projection leads to substantial acceleration for many typical scenarios, where only a single default scattering plane is considered, its benefits are limited for demanding cases involving a two-dimensional grid of scattering angles. Apart from GPU offloading, these cases may benefit from more sophisticated algorithms, such as two-dimensional FFTs applied to successive slices of the polarization distribution to efficiently evaluate the scattered field on a regular grid of transverse wave-vector components~\cite{CHAUMET2015}. Alternative ideas (to be explored) include spherical-harmonic expansions of the voxel polarizations and application of nonequispaced (nonuniform) FFT directly to Eq.~\eqref{eq:far_field_sum}. However, the efficiency and practical benefits of these approaches depend on specific parameters of the scattering problem.

%-------------------------------------------------------------------------------
\bibliographystyle{elsarticle-num}
\bibliography{references}

\begin{thebibliography}{10}
\expandafter\ifx\csname url\endcsname\relax
  \def\url#1{\texttt{#1}}\fi
\expandafter\ifx\csname urlprefix\endcsname\relax\def\urlprefix{URL }\fi
\expandafter\ifx\csname href\endcsname\relax
  \def\href#1#2{#2} \def\path#1{#1}\fi

\bibitem{WRIEDT2009}
T.~Wriedt, Light scattering theories and computer codes, J. Quant. Spectrosc. Radiat. Transfer 110 (2009) 833--843.
\newblock \href {https://doi.org/10.1016/j.jqsrt.2009.02.023} {\path{doi:10.1016/j.jqsrt.2009.02.023}}.

\bibitem{KAHNERT2016}
M.~Kahnert, Numerical solutions of the macroscopic maxwell equations for scattering by non-spherical particles: A tutorial review, J. Quant. Spectrosc. Radiat. Transfer 178 (2016) 22--37.
\newblock \href {https://doi.org/10.1016/j.jqsrt.2015.10.029} {\path{doi:10.1016/j.jqsrt.2015.10.029}}.

\bibitem{PURCELL1973}
E.~M. Purcell, C.~R. Pennypacker, Scattering and absorption of light by nonspherical dielectric grains, Astrophys. J. 186 (1973) 705--714.
\newblock \href {https://doi.org/10.1086/152538} {\path{doi:10.1086/152538}}.

\bibitem{DRAINE1988}
B.~T. Draine, The discrete dipole approximation and its application to interstellar graphite grains, Astrophys. J. 333 (1988) 848--872.
\newblock \href {https://doi.org/10.1086/166795} {\path{doi:10.1086/166795}}.

\bibitem{YURKIN2023}
M.~A. Yurkin, Discrete dipole approximation, in: M.~P. Menguc, M.~Francoeur (Eds.), Light, {Plasmonics} and {Particles}, Elsevier, Amsterdam, 2023, pp. 167--198.
\newblock \href {https://doi.org/10.1016/B978-0-323-99901-4.00020-2} {\path{doi:10.1016/B978-0-323-99901-4.00020-2}}.

\bibitem{GOODMAN1991}
J.~J. Goodman, B.~T. Draine, P.~J. Flatau, Application of fast-fourier-transform techniques to the discrete-dipole approximation, Opt. Lett. 16 (1991) 1198--1200.
\newblock \href {https://doi.org/10.1364/OL.16.001198} {\path{doi:10.1364/OL.16.001198}}.

\bibitem{YURKIN2007}
M.~A. Yurkin, A.~G. Hoekstra, The discrete dipole approximation: an overview and recent developments, J. Quant. Spectrosc. Radiat. Transfer 106 (2007) 558--589.
\newblock \href {https://doi.org/10.1016/j.jqsrt.2007.01.034} {\path{doi:10.1016/j.jqsrt.2007.01.034}}.

\bibitem{CHAUMET2024}
P.~C. Chaumet, A comparative study of efficient iterative solvers for the discrete dipole approximation, J. Quant. Spectrosc. Radiat. Transfer 312 (2024) 108816.
\newblock \href {https://doi.org/10.1016/j.jqsrt.2023.108816} {\path{doi:10.1016/j.jqsrt.2023.108816}}.

\bibitem{SOGABE2022}
T.~Sogabe, Krylov subspace methods for linear systems, Springer Series in Computational Mathematics, 2022.
\newblock \href {https://doi.org/10.1007/978-981-19-8532-4} {\path{doi:10.1007/978-981-19-8532-4}}.

\bibitem{GROTH2020}
S.~P. Groth, A.~G. Polimeridis, J.~K. White, Accelerating the discrete dipole approximation via circulant preconditioning, J. Quant. Spectrosc. Radiat. Transfer 240 (2020) 106689.
\newblock \href {https://doi.org/10.1016/j.jqsrt.2019.106689} {\path{doi:10.1016/j.jqsrt.2019.106689}}.

\bibitem{CHAUMET2023}
P.~C. Chaumet, G.~Maire, A.~Sentenac, Accelerating the discrete dipole approximation by initializing with a scalar solution and using a circulant preconditioning, J. Quant. Spectrosc. Radiat. Transfer 298 (2023) 108505.
\newblock \href {https://doi.org/10.1016/j.jqsrt.2023.108505} {\path{doi:10.1016/j.jqsrt.2023.108505}}.

\bibitem{LANIER2026}
S.~Lanier, Learning to precondition: Reinforcement learning enhanced three-level circulant preconditioning for the discrete dipole approximation, J. Quant. Spectrosc. Radiat. Transfer 350 (2026) 109741.
\newblock \href {https://doi.org/10.1016/j.jqsrt.2025.109741} {\path{doi:10.1016/j.jqsrt.2025.109741}}.

\bibitem{INZHEVATKIN2022}
K.~G. Inzhevatkin, M.~A. Yurkin, Uniform-over-size approximation of the internal fields for scatterers with low refractive-index contrast, J. Quant. Spectrosc. Radiat. Transfer 277 (2022) 107965.
\newblock \href {https://doi.org/10.1016/j.jqsrt.2021.107965} {\path{doi:10.1016/j.jqsrt.2021.107965}}.

\bibitem{SMUNEV2015}
D.~A. Smunev, P.~C. Chaumet, M.~A. Yurkin, Rectangular dipoles in the discrete dipole approximation, J. Quant. Spectrosc. Radiat. Transfer 156 (2015) 67--79.
\newblock \href {https://doi.org/10.1016/j.jqsrt.2015.01.019} {\path{doi:10.1016/j.jqsrt.2015.01.019}}.

\bibitem{YURKIN2010}
M.~A. Yurkin, M.~Min, A.~G. Hoekstra, Application of the discrete dipole approximation to very large refractive indices: {Filtered} coupled dipoles revived, Phys. Rev. E 82 (2010) 036703.
\newblock \href {https://doi.org/10.1103/PhysRevE.82.036703} {\path{doi:10.1103/PhysRevE.82.036703}}.

\bibitem{CHAUMET2004}
P.~C. Chaumet, A.~Sentenac, A.~Rahmani, Coupled dipole method for scatterers with large permittivity, Phys. Rev. E 70 (2004) 036606.
\newblock \href {https://doi.org/10.1103/PhysRevE.70.036606} {\path{doi:10.1103/PhysRevE.70.036606}}.

\bibitem{PILLER1997}
N.~B. Piller, Influence of the edge meshes on the accuracy of the coupled-dipole approximation, Opt. Lett. 22 (1997) 1674--1676.
\newblock \href {https://doi.org/10.1364/OL.22.001674} {\path{doi:10.1364/OL.22.001674}}.

\bibitem{YURKIN2006}
M.~A. Yurkin, V.~P. Maltsev, A.~G. Hoekstra, Convergence of the discrete dipole approximation. ii. an extrapolation technique to increase the accuracy, J. Opt. Soc. Am. A 23 (2006) 2592--2601.
\newblock \href {https://doi.org/10.1364/JOSAA.23.002592} {\path{doi:10.1364/JOSAA.23.002592}}.

\bibitem{PENTTILA2011}
A.~Penttilä, K.~Lumme, Optimal cubature on the sphere and other orientation averaging schemes, J. Quant. Spectrosc. Radiat. Transfer 112 (2011) 1741--1746.
\newblock \href {https://doi.org/10.1016/j.jqsrt.2011.02.001} {\path{doi:10.1016/j.jqsrt.2011.02.001}}.

\bibitem{HOEKSTRA1998}
A.~G. Hoekstra, M.~D. Grimminck, P.~M.~A. Sloot, Large scale simulations of elastic light scattering by a fast discrete dipole approximation, Int. J. Mod. Phys. C 9 (1998) 87--102.
\newblock \href {https://doi.org/10.1142/S012918319800008X} {\path{doi:10.1142/S012918319800008X}}.

\bibitem{HUNTEMANN2011}
M.~Huntemann, G.~Heygster, G.~Hong, Discrete dipole approximation simulations on gpus using opencl—application on cloud ice particles, J. Comput. Sci. 2 (2011) 262--271.
\newblock \href {https://doi.org/10.1016/j.jocs.2011.05.011} {\path{doi:10.1016/j.jocs.2011.05.011}}.

\bibitem{WATERMAN1965}
P.~C. Waterman, Matrix formulation of electromagnetic scattering, Proc. IEEE 53 (1965) 805--812.

\bibitem{WATERMAN1971}
P.~C. Waterman, Symmetry, unitarity, and geometry in electromagnetic scattering, Phys. Rev. D 3 (1971) 825--839.
\newblock \href {https://doi.org/10.1103/PhysRevD.3.825} {\path{doi:10.1103/PhysRevD.3.825}}.

\bibitem{MISHCHENKO2020}
M.~I. Mishchenko, Comprehensive thematic t-matrix reference database: a 2017–2019 update, J. Quant. Spectrosc. Radiat. Transfer 242 (2020) 106692.
\newblock \href {https://doi.org/10.1016/j.jqsrt.2019.106692} {\path{doi:10.1016/j.jqsrt.2019.106692}}.

\bibitem{PETROV2006}
D.~Petrov, E.~Synelnyk, Y.~Shkuratov, G.~Videen, The t-matrix technique for calculations of scattering properties of ensembles of randomly oriented particles with different size, J. Quant. Spectrosc. Radiat. Transfer 102 (2006) 85--110.
\newblock \href {https://doi.org/10.1016/j.jqsrt.2006.02.077} {\path{doi:10.1016/j.jqsrt.2006.02.077}}.

\bibitem{MUINONEN2006}
K.~Muinonen, E.~Zubko, Optimizing the discrete-dipole approximation for sequences of scatterers with identical shapes but differing sizes or refractive indices, J. Quant. Spectrosc. Radiat. Transfer 100 (2006) 288--294.
\newblock \href {https://doi.org/10.1016/j.jqsrt.2005.11.045} {\path{doi:10.1016/j.jqsrt.2005.11.045}}.

\bibitem{OKADA2008}
Y.~Okada, I.~Mann, I.~Sano, S.~Mukai, Acceleration of the iterative solver in the discrete dipole approximation: Application to the orientation variation of irregularly shaped particles, J. Quant. Spectrosc. Radiat. Transfer 109 (2008) 1461--1473.
\newblock \href {https://doi.org/10.1016/j.jqsrt.2008.01.004} {\path{doi:10.1016/j.jqsrt.2008.01.004}}.

\bibitem{UM2013}
E.~S. Um, M.~Commer, G.~A. Newman, Efficient pre-conditioned iterative solution strategies for the electromagnetic diffusion in the earth: finite-element frequency-domain approach, Geophys. J. Int. 193 (2013) 1460--1473.
\newblock \href {https://doi.org/10.1093/gji/ggt071} {\path{doi:10.1093/gji/ggt071}}.

\bibitem{PARKS2006}
M.~L. Parks, E.~De~Sturler, G.~Mackey, D.~D. Johnson, S.~Maiti, Recycling krylov subspaces for sequences of linear systems, SIAM J. Sci. Comput. 28 (2006) 1651--1674.
\newblock \href {https://doi.org/10.1137/040607277} {\path{doi:10.1137/040607277}}.

\bibitem{SOODHALTER2020}
K.~M. Soodhalter, E.~de~Sturler, M.~E. Kilmer, A survey of subspace recycling iterative methods, GAMM-Mitt. 43 (2020) e202000016.
\newblock \href {https://doi.org/10.1002/gamm.202000016} {\path{doi:10.1002/gamm.202000016}}.

\bibitem{OLEARY1980}
D.~P. O'Leary, The block conjugate gradient algorithm and related methods, Linear Algebra Appl. 29 (1980) 293--322.
\newblock \href {https://doi.org/10.1016/0024-3795(80)90247-5} {\path{doi:10.1016/0024-3795(80)90247-5}}.

\bibitem{MOTEKI2026}
N.~Moteki, Block-krylov discrete-dipole and volume-integral-equation solvers for multi-orientation light scattering across the dielectric–metal contrast range, Journal of Quantitative Spectroscopy and Radiative Transfer 362 (2026) 110046.
\newblock \href {https://doi.org/10.1016/j.jqsrt.2026.110046} {\path{doi:10.1016/j.jqsrt.2026.110046}}.

\bibitem{FREUND1993}
R.~W. Freund, Solution of shifted linear systems by quasi-minimal residual iterations, in: L.~Reichel, A.~Ruttan, R.~S. Varga (Eds.), Numerical Linear Algebra, De Gruyter, 1993, pp. 101--122.
\newblock \href {https://doi.org/10.1515/9783110857658.101} {\path{doi:10.1515/9783110857658.101}}.

\bibitem{BUDKO2004}
N.~V. {Budko}, R.~F. {Remis}, {Electromagnetic inversion using a reduced-order three-dimensional homogeneous model}, Inverse Probl. 20 (2004) S17--S26.
\newblock \href {https://doi.org/10.1088/0266-5611/20/6/S02} {\path{doi:10.1088/0266-5611/20/6/S02}}.

\bibitem{YURKIN2011}
M.~A. Yurkin, A.~G. Hoekstra, The discrete-dipole-approximation code {ADDA}: capabilities and known limitations, J. Quant. Spectrosc. Radiat. Transfer 112 (2011) 2234--2247.
\newblock \href {https://doi.org/10.1016/j.jqsrt.2011.01.031} {\path{doi:10.1016/j.jqsrt.2011.01.031}}.

\bibitem{FLETCHER1976}
R.~Fletcher, Conjugate gradient methods for indefinite systems, in: G.~A. Watson (Ed.), Numerical Analysis, Springer, Berlin, 1976, pp. 73--89.
\newblock \href {https://doi.org/10.1007/BFb0080116} {\path{doi:10.1007/BFb0080116}}.

\bibitem{FREUND1992}
R.~W. Freund, Conjugate gradient-type methods for linear systems with complex symmetric coefficient matrices, SIAM J. Sci. Comput. 13 (1992) 425--448.
\newblock \href {https://doi.org/10.1137/0913023} {\path{doi:10.1137/0913023}}.

\bibitem{ARGENTIN2026}
C.~Argentin, P.~C. Chaumet, M.~Gross, M.~A. Yurkin, Floating-point–consistent cross-verification methodology for reproducible and interoperable dda solvers with fair benchmarking, Comput. Phys. Commun. 325 (2026) 110172.
\newblock \href {https://doi.org/10.1016/j.cpc.2026.110172} {\path{doi:10.1016/j.cpc.2026.110172}}.

\bibitem{GUTKOWICZ2004}
D.~Gutkowicz-Krusin, B.~T. Draine, Propagation of electromagnetic waves on a rectangular lattice of polarizable points (2004).
\newblock \href {http://arxiv.org/abs/astro-ph/0403082} {\path{arXiv:astro-ph/0403082}}.

\bibitem{JEGERLEHNER1996}
B.~Jegerlehner, Krylov space solvers for shifted linear systems (1996).
\newblock \href {http://arxiv.org/abs/hep-lat/9612014} {\path{arXiv:hep-lat/9612014}}.

\bibitem{CRAVEN1969}
B.~D. Craven, Complex symmetric matrices, J. Aust. Math. Soc. 10 (1969) 341–354.
\newblock \href {https://doi.org/10.1017/S1446788700007588} {\path{doi:10.1017/S1446788700007588}}.

\bibitem{VORST1990}
H.~van~der Vorst, J.~Melissen, A petrov-galerkin type method for solving axk=b, where a is symmetric complex, IEEE Trans. Magn. 26 (1990) 706--708.
\newblock \href {https://doi.org/10.1109/20.106415} {\path{doi:10.1109/20.106415}}.

\bibitem{HESTENES1952}
M.~Hestenes, E.~Stiefel, Methods of conjugate gradients for solving linear systems, J. Res. Natl. Bur. Stand. 49 (1952) 409--436.
\newblock \href {https://doi.org/10.6028/jres.049.044} {\path{doi:10.6028/jres.049.044}}.

\bibitem{FROMMER1999}
A.~Frommer, P.~Maass, Fast cg-based methods for tikhonov--phillips regularization, SIAM J. Sci. Comput. 20 (1999) 1831--1850.
\newblock \href {https://doi.org/10.1137/S1064827596313310} {\path{doi:10.1137/S1064827596313310}}.

\bibitem{BIRK2015}
S.~Birk, \href{https://elekpub.bib.uni-wuppertal.de/urn/urn:nbn:de:hbz:468-20150601-100922-3}{Deflated shifted block krylov subspace methods for hermitian positive definite matrices}, Ph.D. thesis, University of Wuppertal (2015).
\newline\urlprefix\url{https://elekpub.bib.uni-wuppertal.de/urn/urn:nbn:de:hbz:468-20150601-100922-3}

\bibitem{BARRETT1994}
R.~Barrett, M.~Berry, T.~F. Chan, J.~Demmel, J.~Donato, J.~Dongarra, V.~Eijkhout, R.~Pozo, C.~Romine, H.~van~der Vorst, Templates for the Solution of Linear Systems: Building Blocks for Iterative Methods, SIAM, Philadelphia, 1994.
\newblock \href {https://doi.org/10.1137/1.9781611971538} {\path{doi:10.1137/1.9781611971538}}.

\bibitem{SAAD2003}
Y.~Saad, Iterative Methods for Sparse Linear Systems, 2nd Edition, SIAM, 2003.
\newblock \href {https://doi.org/10.1137/1.9780898718003} {\path{doi:10.1137/1.9780898718003}}.

\bibitem{GREENBAUM1997}
A.~Greenbaum, Estimating the attainable accuracy of recursively computed residual methods, SIAM J. Matrix Anal. A. 18 (1997) 535--551.
\newblock \href {https://doi.org/10.1137/S0895479895284944} {\path{doi:10.1137/S0895479895284944}}.

\bibitem{GUTKNECHT2000}
M.~H. Gutknecht, Z.~Strakos, Accuracy of two three-term and three two-term recurrences for krylov space solvers, SIAM J. Matrix Anal. A. 22 (2000) 213--229.
\newblock \href {https://doi.org/10.1137/S0895479897331862} {\path{doi:10.1137/S0895479897331862}}.

\bibitem{YURKIN2013}
M.~A. Yurkin, Computational approaches for plasmonics, in: F.~Della~Sala, S.~D'Agostino (Eds.), Handbook of {Molecular} {Plasmonics}, Pan Stanford Publishing, Singapore, 2013, pp. 83--138.
\newblock \href {https://doi.org/10.1201/b15328-6} {\path{doi:10.1201/b15328-6}}.

\bibitem{CHAUMET2015}
P.~C. Chaumet, T.~Zhang, A.~Sentenac, Fast far-field calculation in the discrete dipole approximation, J. Quant. Spectrosc. Radiat. Transfer 165 (2015) 88--92.
\newblock \href {https://doi.org/10.1016/j.jqsrt.2015.06.024} {\path{doi:10.1016/j.jqsrt.2015.06.024}}.

\end{thebibliography}

\end{document}